\documentclass[aps,prl,reprint,superscriptaddress,nofootinbib,longbibliography,floatfix]{revtex4-2}
\usepackage[T1]{fontenc}
\usepackage{amsmath,amssymb}
\usepackage{bm}
\usepackage{graphicx}
\usepackage{booktabs}
\usepackage{xcolor}
\usepackage{textcomp}
\usepackage[colorlinks=true,linkcolor=blue!50!black,citecolor=blue!50!black,urlcolor=blue!50!black]{hyperref}

\newcommand{\methodsec}[1]{\par\medskip\noindent\textbf{#1}\par\nobreak\noindent\ignorespaces}
\newcommand{\methodsub}[1]{\par\smallskip\noindent\textit{#1}\par\nobreak\noindent\ignorespaces}
\newcommand{\iswap}[1]{\mathrm{iSWAP}^{#1}}
\newcommand{\XYP}{\operatorname{XY2P}}
\begin{document}

\title{Imaging chiral snake states in a qubit array with programmable gauge fields}

\author{Hanyi Wang}
\thanks{These authors contributed equally to this work.}
\affiliation{China Telecom Quantum Information Technology Group Co., Ltd., Hefei 230094, Anhui, China}
\affiliation{Quantum Research Institute of China Telecom, Hefei 230031, Anhui, China}
\author{Lijun Sun}
\thanks{These authors contributed equally to this work.}
\affiliation{China Telecom Quantum Information Technology Group Co., Ltd., Hefei 230094, Anhui, China}
\affiliation{Quantum Research Institute of China Telecom, Hefei 230031, Anhui, China}
\author{Bihao Guo}
\affiliation{China Telecom Quantum Information Technology Group Co., Ltd., Hefei 230094, Anhui, China}
\affiliation{Quantum Research Institute of China Telecom, Hefei 230031, Anhui, China}
\author{Weizhi Tao}
\affiliation{China Telecom Quantum Information Technology Group Co., Ltd., Hefei 230094, Anhui, China}
\affiliation{Quantum Research Institute of China Telecom, Hefei 230031, Anhui, China}
\author{Jian Ding}
\affiliation{China Telecom Quantum Information Technology Group Co., Ltd., Hefei 230094, Anhui, China}
\affiliation{Quantum Research Institute of China Telecom, Hefei 230031, Anhui, China}
\author{Shengbin Wang}
\affiliation{China Telecom Quantum Information Technology Group Co., Ltd., Hefei 230094, Anhui, China}
\affiliation{Quantum Research Institute of China Telecom, Hefei 230031, Anhui, China}
\author{Xinfang Zhang}
\affiliation{China Telecom Quantum Information Technology Group Co., Ltd., Hefei 230094, Anhui, China}
\affiliation{Quantum Research Institute of China Telecom, Hefei 230031, Anhui, China}
\author{Wei-Yong Zhang}
\affiliation{Hefei National Research Center for Physical Sciences at the Microscale and School of Physical Sciences, University of Science and Technology of China, Hefei 230026, China}
\affiliation{Hefei National Laboratory, University of Science and Technology of China, Hefei 230088, China}
\author{Hantao Sun}

\affiliation{China Telecom Quantum Information Technology Group Co., Ltd., Hefei 230094, Anhui, China}
\affiliation{Quantum Research Institute of China Telecom, Hefei 230031, Anhui, China}

\begin{abstract}

Spatially varying magnetic fields steer charged particles along directed paths. 
Where a field reverses sign, particles follow snake-like orbits along the reversal line, in a direction set by both the field orientation and the sign of their energy~\cite{Muller1992electron,Reijniers2000snake}. 
Transport measurements have detected such snake states~\cite{Nogaret2000resistance,Williams2011snake,Rickhaus2015snake,Taychatanapat2015conductance}, but imaging their dynamics in a reprogrammable field landscape remains challenging. 
Here we digitally simulate the Harper--Hofstadter model with programmable synthetic magnetic fields~\cite{Dalibard2011colloquium,Ozawa2019topological} on a 49-qubit superconducting processor. 
In each Trotter step, every hopping term is one native fractional-iSWAP gate, and single-qubit $Z$ rotations set all 36 plaquette fluxes independently, without additional entangling gates. 
Uniform, random, and stepped flux landscapes reveal Aharonov--Bohm suppression of diagonal propagation, disorder-induced slowing, and magnetic reflection. 
At a field reversal, bonding and antibonding initial states, which carry opposite energy biases, propagate in opposite directions, and both directions reverse when the field is inverted. 
A symmetry analysis of these reversals yields a chiral displacement of nearly one lattice spacing after ten Trotter rounds, with symmetry-breaking residuals below 0.05 lattice spacings. 
Bond-resolved flux programming with site-resolved imaging thus provides a quantitative tool for synthetic gauge fields and a route to time-dependent gauge fields and interacting particles.

\end{abstract}

\maketitle

A charged particle that moves around a closed path acquires a phase proportional to the enclosed magnetic flux~\cite{Aharonov1959significance}. 
On a lattice, this phase enters the hopping amplitudes through the Peierls substitution~\cite{Peierls1933theorie}, and a flux per plaquette comparable to a flux quantum gives rise to the Hofstadter spectrum, whose bands can carry nonzero Chern numbers~\cite{Harper1955single,Azbel1964energy, Hofstadter1976energy,Thouless1982quantized}. 
Reaching this regime in ordinary crystals would require magnetic fields far beyond laboratory capabilities. Superlattices~\cite{Dean2013hofstadter}, mesoscopic networks~\cite{Pannetier1984experimental} and synthetic gauge fields~\cite{Dalibard2011colloquium,Ozawa2019topological} in ultracold atoms~\cite{Aidelsburger2013realization,Miyake2013realizing,Aidelsburger2015measuring,Stuhl2015visualizing,Leonard2023realization}, photonic resonators~\cite{Hafezi2013imaging} and superconducting circuits~\cite{Roushan2017chiral,Owens2022chiral,Rosen2024synthetic,Wang2024fractional}
have brought it within reach, almost always with a uniform flux. 
Exploring magnetic disorder and field boundaries requires a further capability: control of the flux from one plaquette to the next. 
For example, random flux modifies localization and models the emergent gauge fields of correlated matter~\cite{Furusaki1999anderson,Evers2008anderson,Halperin1993theory,LeeNagaosa1992gauge}, while a step in the flux can reflect or guide a particle~\cite{Matulis1994wavevector,An2017direct}. 
Towards this goal, programmable flux textures have recently been realized in a momentum-space lattice~\cite{Wang2026programmable}, where each control phase is shared by several links.

Among these spatial flux textures, a field-reversal interface provides a direct setting for probing chiral transport. 
Cyclotron orbits curve in opposite senses on its two sides and join into snake-like paths that run along the line~\cite{Muller1992electron,Reijniers2000snake}. 
Classically, the field orientation fixes the direction of this motion. 
Quantum mechanically, the direction also depends on the sign of the energy~\cite{Reijniers2000snake,Oroszlany2008theory}. Reversing the field and reversing the sign of the energy are therefore two independent ways of reversing the path. 
Snake states have been detected in transport measurements on semiconductor and graphene devices~\cite{Nogaret2000resistance,Williams2011snake,Rickhaus2015snake,Taychatanapat2015conductance}, but these measurements do not directly image the propagation of a wave packet with a chosen energy bias through a reprogrammable field landscape. 
Imaging these dynamics requires control of both the initial wave packet and the field landscape, together with site-resolved detection.

\begin{figure*}[t]
    \centering
    \includegraphics[width=\textwidth]{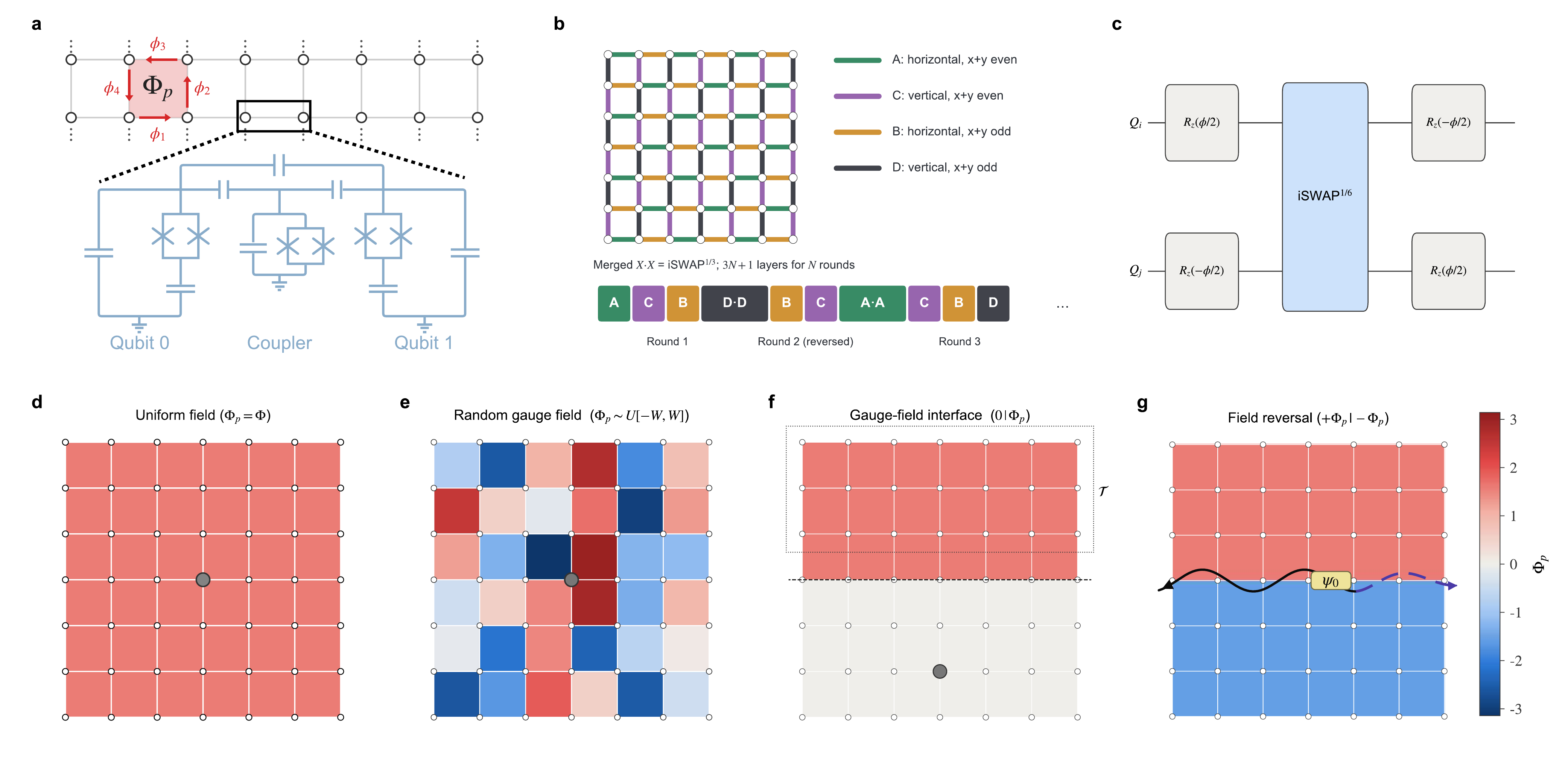}
    \caption{\textbf{Programming a synthetic magnetic field bond by bond.} 
    \textbf{a}, A $7\times7$ array of 49 frequency-tunable transmon qubits connected by tunable couplers (bottom: circuit diagram of one qubit--coupler--qubit unit). 
    The synthetic flux $\Phi_p$ through each plaquette is the oriented sum of the Peierls phases $\phi_{ij}$ on its four bonds. 
    \textbf{b}, The 84 bonds form four groups of disjoint bonds (A--D), 21 per group. 
    A Trotter round applies the four layers in the order A, C, B, D; the next round reverses the order, so that the identical layers at each round boundary merge into a single layer of $\iswap{1/3}$ gates.
    \textbf{c}, Phase-programmed exchange gate: single-qubit $R_z$ rotations before and after a native $\iswap{1/6}$ gate (transfer angle $\theta=\pi/12$) set the hopping phase $\phi$ and compensate the dynamical phases of the gate. 
    \textbf{d--g}, Programmed flux landscapes: \textbf{d}, uniform field; 
    \textbf{e}, zero-mean random field; 
    \textbf{f}, $0\,|\,\Phi$ interface (zero flux on the lower three plaquette rows, $\Phi$ on the upper three), with $\mathcal T$ the transmitted region; 
    \textbf{g}, $+\Phi\,|\,-\Phi$ reversal along the central row ($+\Phi$ above, $-\Phi$ below), with the two-site initial state $\psi_0$ and schematic snake orbits for the two signs of the energy (solid and dashed), which run in opposite directions.
    Dots mark single-site initial states; the colour bar gives the plaquette flux.}
    \label{fig:concept}
\end{figure*}

Here we implement the Harper--Hofstadter (HH) model digitally on a $7\times7$ superconducting-qubit array (Fig.~\ref{fig:concept}).
In each Trotter step, every hopping term is a single native fractional-iSWAP gate~\cite{Foxen2020demonstrating,Wang2026lifting}, rather than the continuous exchange of analogue quantum walks ~\cite{Yan2019strongly,Gong2021quantum}, and single-qubit $Z$ rotations set its Peierls phase. 
All 36 plaquette fluxes are therefore programmed independently, without additional entangling gates. 
We first validate this control with uniform, random, and stepped fields, through Aharonov--Bohm interference, disorder-induced slowing, and magnetic reflection.  
We then image bonding and antibonding initial states propagating in opposite directions along a field reversal and reverse both trajectories by inverting the field. 
Using the symmetry of the model, we extract a chiral displacement that is insensitive at first order to static on-site detunings and phase offsets shared by all four state--field configurations, together with two residuals that quantify symmetry breaking. 
A spatial scan of the initial state separates the interface channel from oppositely directed edge transport. 
Together, these measurements provide a quantitative, site-resolved test of chiral transport in a programmable gauge field.

\section{A gauge field written bond by bond}

We encode the HH model in the single-excitation manifold of a $7\times7$ array of frequency-tunable transmon qubits connected by tunable couplers, based on the architecture of ref.~\cite{Gao2025establishing} (Fig.~\ref{fig:concept}a; device and gate benchmarks in Extended Data Fig.~\ref{fig:device} and Extended Data Table~\ref{tab:device}).
Within this manifold, the array emulates the Hamiltonian

\begin{equation}
    \begin{aligned}
    H &= -J\sum_{\langle i,j\rangle}
    \left(e^{i\phi_{ij}}a_j^{\dagger}a_i
    +\mathrm{h.c.}\right),\\
    \Phi_p &= \sum_{\langle i,j\rangle\in\partial p}
    \phi_{ij}\pmod{2\pi},
    \end{aligned}
    \label{eq:H}
\end{equation}
where $a_i^{\dagger}$ creates an excitation on qubit $i$, $\phi_{ij}$ is the Peierls phase of the oriented bond from $i$ to $j$, $J$ is the hopping amplitude and $\hbar=1$. 
The second sum runs around the boundary $\partial p$ of plaquette $p$; with bonds oriented along $+x$ and $+y$, it equals $\phi_{\rm bottom}+\phi_{\rm right}-\phi_{\rm top}-\phi_{\rm left}$ (Methods). 
We approximate $e^{-iHt}$ by a Trotter--Suzuki decomposition~\cite{Lloyd1996universal,Suzuki1990fractal}. 
The 84 nearest-neighbour bonds are split into four groups of 21 disjoint bonds (A--D in Fig.~\ref{fig:concept}b). 
Each hopping term is one native $\iswap{1/6}$ gate~\cite{Wang2026lifting} with transfer angle $\theta=J\delta t=\pi/12$, and all gates in a group run in parallel. 
The layer order alternates between rounds, which makes each pair of rounds a symmetric second-order product formula. 
Identical layers at each round boundary merge into $\iswap{1/3}$ gates, so $N$ rounds require $3N+1$ exchange layers (Methods and Extended Data Fig.~\ref{fig:trotter}).

\begin{figure*}[t]
    \centering
    \includegraphics[width=\textwidth]{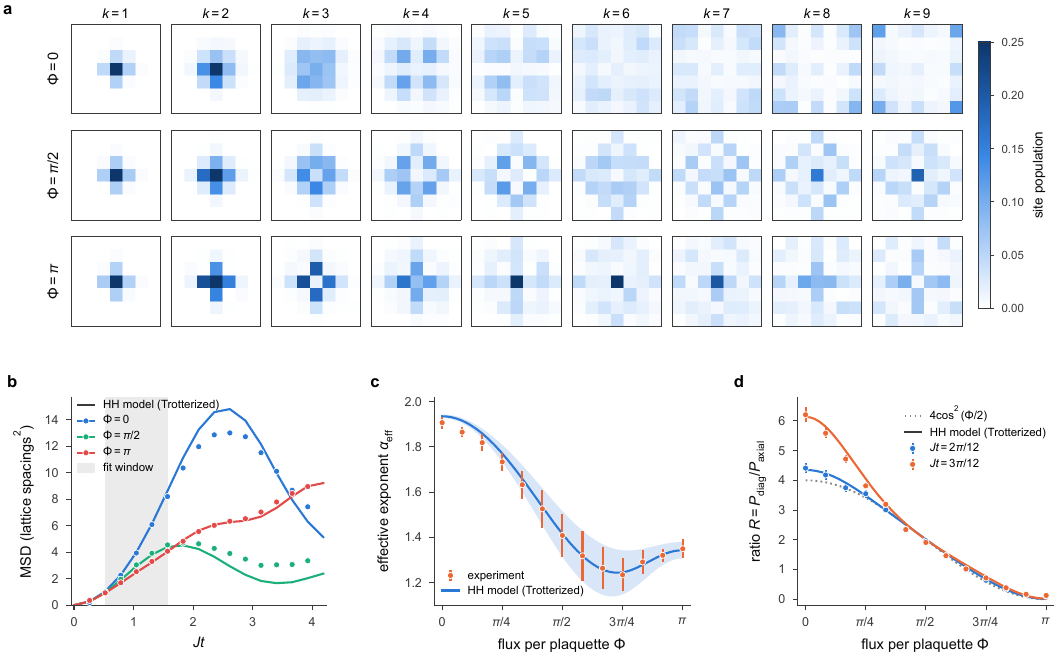}
    \caption{\textbf{Aharonov--Bohm interference and transport in a uniform synthetic magnetic field.}
    \textbf{a}, Measured site populations after Trotter rounds $k=1$--$9$ ($Jt=k\pi/12$) for $\Phi=0$, $\pi/2$ and $\pi$. 
    Colour scale, site population (saturates at 0.25). 
    Diagonal propagation is suppressed at $\Phi=\pi$, whereas the wave packet partially refocuses onto the initial site after eight to nine rounds at $\Phi=\pi/2$. 
    Predictions of the Harper--Hofstadter (HH) model for both Trotterized and continuous-time evolution are shown in Extended Data Fig.~\ref{fig:density}.
    \textbf{b}, Mean-square displacement (MSD) versus $Jt$ for the same three fluxes. 
    Shaded region, fit window (rounds 2--6) used to extract the effective transport exponent from $\mathrm{MSD}\propto t^{\alpha_{\rm eff}}$. 
    \textbf{c}, Effective exponent $\alpha_{\rm eff}$ versus flux for thirteen programmed values. 
    Shaded band, $\pm$ one standard error (from the fit residuals) of the exponent fitted to the Trotterized-model MSD over the same window. 
    \textbf{d}, Interference ratio $R=P_{\rm diag}/P_{\rm axial}$ of the diagonal to the axial second-neighbour populations after two and three rounds. 
    Dotted line, short-time prediction $4\cos^2(\Phi/2)$. 
    In \textbf{b--d}, symbols show experimental data and solid lines show the Trotterized HH model. Error bars denote one standard error: in \textbf{b} and \textbf{d}, counting uncertainties propagated through the readout correction; in \textbf{c}, uncertainties of the fitted exponents (Methods). Error bars in \textbf{b} are smaller than the symbols.}
    \label{fig:uniform}
\end{figure*}

We encode each Peierls phase with single-qubit $R_z$ rotations before and after the exchange gate of its bond (Fig.~\ref{fig:concept}c). 
The same rotations compensate for the calibrated dynamical phases of the native interaction. 
A $Z$ rotation on one qubit does not commute with the fractional iSWAP, so these rotations cannot simply be deferred or absorbed into later gates. 
We apply them as physical pulses and reduce their number by merging adjacent rotations and by bit-flip frames (Methods). 
All 36 plaquette fluxes can thus be programmed independently with the same calibrated exchange gates and without additional entangling gates.

Because the target dynamics conserve the excitation number, we measure all 49 qubits after each evolution and retain only outcomes with exactly one excitation; we then correct readout errors within the single-excitation manifold (Methods). 
Each circuit yields more than 4,000 post-selected outcomes, and all error bars denote one standard error. 
For the uniform-field circuits, the post-selection rate falls from 43–48\% after one round to 1.5–2.0\% after fifteen rounds (46 exchange layers), consistent with the benchmarked gate errors. 
Shots that have gained or lost an excitation can still pass post-selection through a compensating readout or relaxation error; an error-budget model constrained by the measured excitation-number statistics estimates this residual background at about 5\%, 12\% and 20\% of the retained shots after four, eight and twelve rounds (Methods). 
Throughout, we compare the data with exact simulations of the ideal, noiseless Trotter circuit, which we refer to below as the Trotterized HH model.

\section{A uniform synthetic field}

A uniform field provides the first test of the flux programming. 
We program the same flux $\Phi$ on all 36 plaquettes and initialize the excitation at the central site (Fig.~\ref{fig:uniform}a). 
At $\Phi=0$, the wavefront expands ballistically and reaches the corners by the ninth round. 
At $\Phi=\pi/2$, it spreads into a ring and partially refocuses after eight to nine rounds, as expected for cyclotron motion on a lattice. 
At $\Phi=\pi$, propagation is concentrated along the two lattice axes through the initial site. 
The Trotterized and continuous-time HH dynamics show the same patterns, with total-variation distances of at most 0.15 between their site populations at $Jt=\pi/2$ (Extended Data Figs.~\ref{fig:density} and~\ref{fig:trotter}a).

The mean-square displacement, $\mathrm{MSD}(t)=\sum_i|\mathbf r_i-\mathbf r_0|^2P_i(t)$, measures the spreading (Fig.~\ref{fig:uniform}b). 
For the target Harper--Hofstadter Hamiltonian, the short-time expansion (with $\hbar=1$ and unit lattice spacing) is
\begin{equation}
    \mathrm{MSD}(t)=4J^2t^2-\frac{J^4t^4}{3}
    \sum_{p\ni\mathbf r_0}\bigl(1-\cos\Phi_p\bigr)+O(t^6),
    \label{eq:msd}
\end{equation}
where the sum runs over the four plaquettes surrounding the initial site. 
The leading term describes ballistic spreading and is independent of the flux. 
The flux enters at fourth order in $t$, through interference between the two two-step paths around each plaquette adjacent to the initial site, and slows the spreading.

\begin{figure*}[t]
    \centering
    \includegraphics[width=\textwidth]{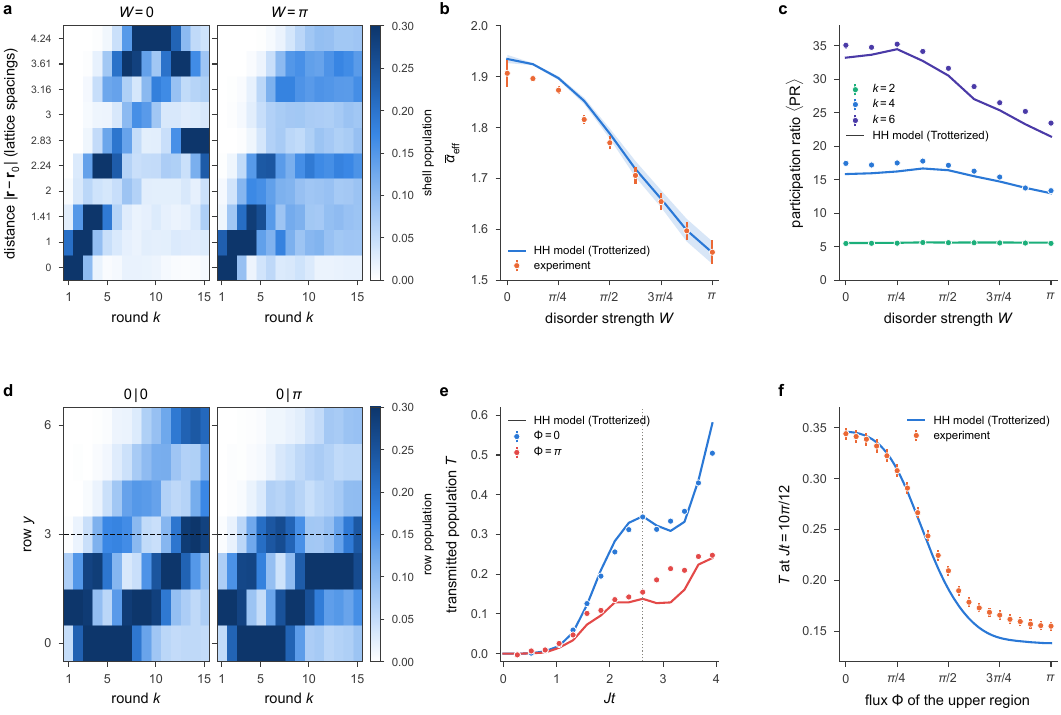} 
    \caption{\textbf{Random flux disorder and magnetic reflection.} 
    \textbf{a}, Measured shell population (sum over sites at equal distance from the initial site) for $W=0$ and $W=\pi$ versus Trotter round $k=1$--$15$ ($Jt=k\pi/12$). 
    Plaquette fluxes are drawn independently from a uniform distribution on $[-W,W]$. 
    Disorder averages in \textbf{a}--\textbf{c} use 20 realizations for each nonzero $W$. 
    \textbf{b}, Mean effective transport exponent $\overline{\alpha}_{\rm eff}$ versus $W$. 
    For $W>0$, the MSD is fitted over rounds 2--6 for each of 20 disorder realizations, and the resulting exponents are averaged. 
    Shading denotes $\pm$ one standard error of the mean across the 20 model exponents. 
    At $W=0$, data, model, error bar, and shading are those of Fig.~\ref{fig:uniform}c (no disorder averaging).
    \textbf{c}, Realization-averaged participation ratio versus $W$ after 2, 4 and 6 rounds. 
    \textbf{d}, Measured row-resolved populations versus Trotter round for the $0\,|\,0$ and $0\,|\,\pi$ landscapes. 
    Dashed lines, interface row; geometry in Fig.~\ref{fig:concept}f. 
    \textbf{e}, Population $\mathcal T$ transmitted into the upper region versus $Jt$ for $\Phi=0$ and $\pi$. 
    Dotted line, $Jt=10\pi/12$. 
    \textbf{f}, Transmitted population at $Jt=10\pi/12$ versus flux contrast for 21 programmed values. 
    In \textbf{b}, \textbf{c}, \textbf{e} and \textbf{f}, symbols show experimental data and solid lines show the Trotterized Harper--Hofstadter (HH) model, evaluated for the same flux landscapes as the experiment. 
    For $W>0$, experimental error bars in \textbf{b} combine, in quadrature, the propagated fit uncertainties and the standard error across disorder realizations. 
    Error bars in \textbf{c}, \textbf{e} and \textbf{f} represent counting uncertainties propagated through readout correction and the calculation of the plotted observables, including disorder averaging in \textbf{c} (Methods). 
    Where not visible, error bars are smaller than the symbols.}
    \label{fig:random}
\end{figure*}

We fit $\mathrm{MSD}\propto t^{\alpha_{\rm eff}}$ over rounds 2--6, before boundary reflections set in. 
The effective exponent decreases from $1.91\pm0.03$ at $\Phi=0$ to $1.23\pm0.07$ at $\Phi=3\pi/4$, then rises to $1.35\pm0.04$ at $\Phi=\pi$ (Fig.~\ref{fig:uniform}c). 
These values follow the Trotterized HH model, whose exponent is non-monotonic in $\Phi$, with a minimum near $\Phi=0.73\pi$ that reflects the finite fit window rather than asymptotic transport (Methods). 
The measured rise between $3\pi/4$ and $\pi$ is consistent with this dependence but is not resolved by the data alone. 

At $\Phi=\pi$, Aharonov--Bohm interference suppresses diagonal propagation, favouring spreading along the lattice axes. 
Each diagonal neighbour of the initial site is reached by two shortest paths enclosing one plaquette, with a combined amplitude proportional to $1+e^{i\Phi}$. 
By contrast, each second neighbour along an axis is reached by a single shortest path. 
The population ratio $R=P_{\rm diag}/P_{\rm axial}$ therefore approaches $4\cos^2(\Phi/2)$ in the short-time limit.

Measurements after two and three rounds at thirteen flux values show this suppression of diagonal propagation (Fig.~\ref{fig:uniform}d). 
After two rounds, $R$ decreases from $4.4\pm0.2$ at $\Phi=0$ to $0.11\pm0.01$ at $\Phi=\pi$; after three rounds, the corresponding values are $6.2\pm0.2$ and $0.13\pm0.01$. 
These finite-time ratios are consistent with the Trotterized HH model. 
Near $\Phi=\pi$, the measured ratios remain slightly above the expected model values, which tend to approach zero. 
We attribute this discrepancy to residual background that persists after post-selection and populates the nominally dark diagonal sites (Methods).

\section{Random fields and a field interface}

Local flux control allows us to introduce magnetic disorder without changing the calibrated exchange gates. 
We draw each plaquette flux independently from the zero-mean distribution $\Phi_p\sim U[-W,W]$, realizing a lattice random-flux model~\cite{Furusaki1999anderson,Evers2008anderson}. 
For each nonzero disorder strength $W$, we average the site populations over 20 independent realizations. The $W=0$ reference uses the uniform zero-field data. 
As $W$ increases, the disorder-averaged density remains closer to the origin and the coherent revivals of the clean lattice disappear (Fig.~\ref{fig:random}a). 
For each realization we fit the MSD over rounds 2--6 and then average the exponents. 
The mean exponent decreases from $1.91\pm0.03$ at $W=0$ to $1.55\pm0.02$ at $W=\pi$ (Fig.~\ref{fig:random}b), which is in agreement with the Trotterized HH model for the same flux landscapes. 

For this distribution, $\langle\cos\Phi_p\rangle=\sin W/W$ in equation~(\ref{eq:msd}), so the short-time suppression is weaker than for a uniform flux $\Phi=W$. 
To quantify the spatial extent, we use the realization-averaged participation ratio $\langle\mathrm{PR}\rangle=\langle(\sum_iP_i^2)^{-1}\rangle$, the effective number of occupied sites. 
The reduction in participation ratio with increasing $W$ becomes more pronounced at later rounds (Fig.~\ref{fig:random}c). 
After six rounds, $\langle\mathrm{PR}\rangle$ decreases from $35.1\pm0.5$ at $W=0$ to $23.5\pm0.4$ at $W=\pi$. 
Within the accessible time window, the disorder slows the spreading. 
Continuous-time simulations on a $61\times61$ lattice show a local exponent approaching 1 at late simulated times, consistent with diffusive spreading over that interval (Methods and Extended Data Fig.~\ref{fig:randomtheory}).

We next introduce a magnetic interface, denoted $0\,|\,\Phi$, by setting the lower three rows of plaquettes to zero flux and the upper three to $\Phi$ (Fig.~\ref{fig:concept}f). 
The excitation starts two rows below the interface, and we measure the population $\mathcal T(t)$ entering the upper region. 
At $\Phi=0$, the excitation spreads freely across the nominal interface. 
At $\Phi=\pi$, population accumulates near the interface on the incident side, and transmission is suppressed (Fig.~\ref{fig:random}d,e).

After ten rounds ($Jt=10\pi/12$), $\mathcal T$ decreases with increasing flux contrast (Fig.~\ref{fig:random}f). 
It falls from $0.344\pm0.006$ at $\Phi=0$ to $0.209\pm0.005$ at $\Phi=\pi/2$ and $0.155\pm0.005$ at $\Phi=\pi$. 
The corresponding predictions of the Trotterized HH model are $0.346$, $0.191$ and $0.138$. 
The rms deviation between experiment and model is $0.014$ across all 21 flux values, with measurements lying slightly above the predictions at large flux contrast. 
The suppression of transmission indicates that the flux step acts as a magnetic barrier~\cite{Matulis1994wavevector,An2017direct}.

\section{Imaging chiral snake states}

\begin{figure*}[t]
    \centering
    \includegraphics[width=\textwidth]{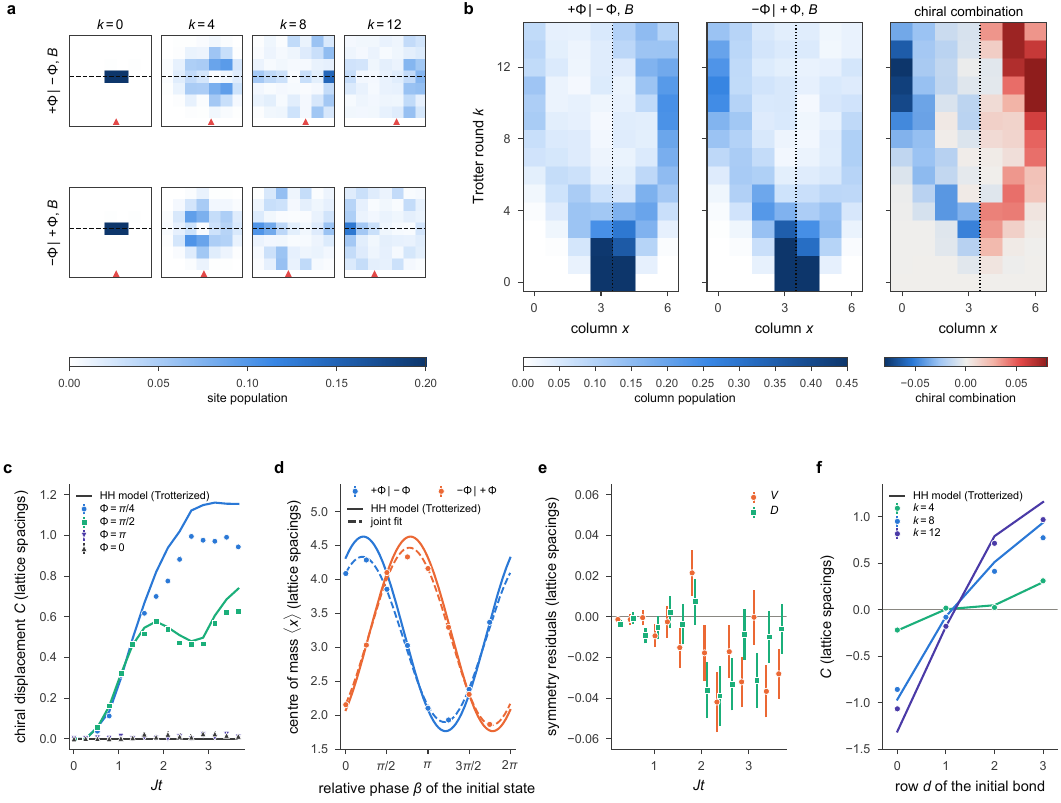} 
    \caption{\textbf{Imaging chiral snake states at a field-reversal interface.} 
    \textbf{a}, Measured site populations of the bonding state after $k=0$, 4, 8 and 12 rounds for $+\Phi\,|\,-\Phi$ (top) and $-\Phi\,|\,+\Phi$ (bottom) at $\Phi=\pi/4$. Dashed lines, reversal row (Fig.~\ref{fig:concept}g); red triangles, horizontal centre-of-mass position at the corresponding round; colour scale, site population (saturates at 0.2).
    \textbf{b}, Measured column populations versus round for the bonding state in both orientations, together with the chiral combination $(P_{+,B}-P_{+,A}-P_{-,B}+P_{-,A})/4$, where $P_{s,\eta}$ is the column population for orientation $s=\pm$ and state $\eta=B,A$. Dotted lines, centre of the initial bond.
    \textbf{c}, Chiral displacement $C$ (equation~(\ref{eq:C})) versus $Jt$ for $\Phi=\pi/4$, $\pi/2$, $\pi$ and 0.
    \textbf{d}, Centre of mass versus the initial relative phase $\beta$ after ten rounds ($Jt=10\pi/12$) for both orientations. Dashed curves, joint weighted fit of the sixteen centroids to equation~(\ref{eq:fit}); solid curves, Trotterized HH model. 
    \textbf{e}, Symmetry residuals $V$ and $D$ (defined in the text) versus $Jt$. Horizontal line, zero.
    \textbf{f}, Chiral displacement after 4, 8 and 12 rounds versus the row $d$ of the initial bond; $d=3$ is the reversal row and $d=0$ the lower edge, where the displacement reverses sign. In \textbf{c--f}, symbols show experimental data; solid curves in \textbf{c}, \textbf{d} and \textbf{f} show the Trotterized Harper--Hofstadter (HH) model and dashed curves in \textbf{d} the fit. Positions and displacements are in lattice spacings. Error bars denote one standard error from counting statistics, propagated through the readout correction to the centres of mass in \textbf{d} and to their four-measurement combinations in \textbf{c}, \textbf{e} and \textbf{f} (Methods); the uncertainty quoted for the chiral amplitude $f$ is one standard error of the fit.}
    \label{fig:chiral}
\end{figure*}

We next reverse the sign of the field across the interface and ask whether a wave packet travels along it. 
A single-site initial state has equal weight at positive and negative energies, and the bipartite symmetry $P(-\Phi,\psi)=P(+\Phi,S\psi^*)$, with $S=\mathrm{diag}[(-1)^{x+y}]$, makes its site populations invariant under field reversal (Methods).
To bias the energy distribution, we prepare bonding and antibonding states on two neighbouring sites. 
In the gauge where the initial bond phase vanishes, $|B\rangle=(|10\rangle+|01\rangle)/\sqrt2$ and $|A\rangle=(|10\rangle-|01\rangle)/\sqrt2$ have mean energies $-J$ and $+J$, so changing only their relative phase selects the sign of the energy bias. 
The symmetry maps $|B\rangle$ onto $|A\rangle$, relating bonding-state propagation in one field orientation to antibonding-state propagation in the other. 
We prepare these states on the central bond with a separately calibrated $\iswap{1/2}$ gate and program opposite fluxes above and below the central row (Fig.~\ref{fig:concept}g).

At $\Phi=\pi/4$, the magnetic length $\ell_B=a/\sqrt{\Phi}\approx1.1a$ places the observed snake-state motion on the lattice scale. 
The field reversal supports interface-localized eigenstates paired by bipartite symmetry at opposite energies and group velocities (Extended Data Fig.~\ref{fig:spectra}). 
The bonding and antibonding initial states carry 74\% of their spectral weight at negative and positive energies, respectively, and drift in opposite directions. 
With $+\Phi$ above the interface and $-\Phi$ below it, the bonding state moves to the right; reversing the field reverses both drift directions. 
The population remains confined within about one row of the reversal line (Fig.~\ref{fig:chiral}a,b), and the drift saturates after about ten rounds, when the population reaches the end of the seven-site interface (Fig.~\ref{fig:chiral}b,c).

To isolate this motion from common drifts, we combine the centres of mass $X_{s,\eta}$ for the two orientations $s=\pm$ and states $\eta=B,A$ into
\begin{equation}
    C=\tfrac14\left(X_{+,B}-X_{+,A}-X_{-,B}+X_{-,A}\right),
    \label{eq:C}
\end{equation}
which changes sign under either reversal. 
At $\Phi=\pi/4$, the chiral displacement reaches $0.99\pm0.01$ lattice spacings after ten rounds, compared with $1.12$ for the Trotterized HH model (Fig.~\ref{fig:chiral}c). 
Over rounds 4--10, the linear growth at $\Phi=\pi/4$ gives a chiral velocity of $0.44\pm0.01\,Ja/\hbar$. 
The same fit gives $0.52\,Ja/\hbar$ for the Trotterized circuit and $0.58\,Ja/\hbar$ for continuous-time evolution on the open array. 
The difference between these two values is the digitization error; the measured velocity shows a reduction comparable to that of the chiral displacement (Methods). 
The control landscapes at $\Phi=0$ and $\pi$ give $|C|\lesssim0.025$ throughout: at $\Phi=0$ there is no interface, and at $\Phi=\pi$ the fluxes $+\pi$ and $-\pi$ are equivalent modulo $2\pi$.

To resolve the phase dependence of the chiral response, we scan the relative phase $\beta$ of the initial state $(|10\rangle+e^{i\beta}|01\rangle)/\sqrt2$ (Fig.~\ref{fig:chiral}d), jointly fitting both field orientations ($s=\pm1$) to
\begin{equation}
X(\beta,s) = a_0 + a_1s + (b+sf)\cos\beta + (g+sh)\sin\beta.
\label{eq:fit}
\end{equation}
This separates the chiral amplitude $f$ from the field-even drift $g$ associated with the initial current. The ideal Trotterized HH model maps $\beta$ to $\pi-\beta$ under field reversal, requiring $a_1=b=h=0$ and hence $\tfrac12[X_+(\beta)-X_-(\beta)]=C\cos\beta$. It predicts no chiral displacement at $\beta=\pm\pi/2$, where the states carry current but have zero mean energy.

At round 10, the fitted chiral amplitude, $f=0.99\pm0.01$, agrees with $C=0.99\pm0.01$ obtained from the four state--field combinations, and the field-even drift is $g=0.79\pm0.01$. 
Two coefficients that should vanish in the ideal model are small but nonzero, $h=-0.06\pm0.01$ and $a_1=-0.02\pm0.01$, and they quantify the departure from field-reversal symmetry (Methods).

The four measurements used to construct $C$ also yield two symmetry diagnostics,
\[
\begin{aligned}
V&=\tfrac14\left(X_{+,B}+X_{-,B}-X_{+,A}-X_{-,A}\right),\\
D&=\tfrac14\left(X_{+,B}-X_{-,B}+X_{+,A}-X_{-,A}\right),
\end{aligned}
\]
both of which vanish in the ideal Trotterized HH model. 
In the shared static-error model, $V$ and $D$ respond to detunings and phase offsets at first order, whereas corrections to $C$ enter at second order (Methods). 
Experimentally, both remain below $0.045$ in magnitude, with resolved late-time values between $-0.03$ and $-0.04$ (Fig.~\ref{fig:chiral}e).

To distinguish interface transport from edge motion, we move the initial bond from the reversal line towards the lower edge. 
At $\Phi=\pi/4$ after twelve rounds, $C$ decreases to $0.72\pm0.01$ one row below the field-reversal line and $-0.18\pm0.02$ two rows below it. 
At the lower edge, it reaches $-1.07\pm0.02$, consistent with oppositely directed skipping motion (Fig.~\ref{fig:chiral}f). 
The Trotterized HH model captures this spatial dependence, including the crossover from interface snake motion to edge transport in the opposite direction.

\section{Outlook}

In summary, we have programmed a synthetic gauge field bond by bond on a 49-qubit superconducting processor. 
Each hopping term of the Harper--Hofstadter Hamiltonian is one native fractional-iSWAP gate per Trotter step, and its Peierls phase is a single-qubit rotation angle, so uniform, random, and stepped landscapes run on the same calibrated exchange gates: the gauge field becomes a circuit parameter. 
Aharonov--Bohm interference, disorder-induced slowing, and magnetic reflection validate this control against the Trotterized HH model. 
At a field reversal, bonding and antibonding wave packets show energy-selective snake-state transport directly in space and time. 
Under the shared static-error model, the four reversal measurements separate out a chiral displacement whose first-order correction cancels from two residuals that can respond at first order to symmetry breaking.
Exact calculations of the single-excitation dynamics allow us to quantify deviations from the ideal circuit and distinguish them from digitization effects (Methods).

Bond-resolved control also points to further extensions.
Updating rotation angles between Trotter rounds would generate time-dependent gauge fields, including synthetic electric fields, without additional entangling gates. 
Preparing two excitations at the field-reversal interface would give access to hard-core boson transport. 
Combining a tunable $ZZ$ component with fractional iSWAP exchange in a single native two-qubit gate could introduce tunable nearest-neighbour density interactions~\cite{Wang2026lifting}. 
With local phase shifts compensated, this extension would preserve the Peierls-phase encoding and entangling-gate count, enabling studies of density-dependent transport along the interface. 
Preparing correlated many-body states at fractional particle-to-flux filling could open a route to fractional quantum Hall physics~\cite{Sorensen2005fractional,Leonard2023realization,Wang2024fractional}. 
The practical limits are circuit depth and array size: the post-selection rate falls with the number of Trotter rounds, and the $7\times7$ array limits the interface to seven sites. As these limits are relaxed, independent plaquette control and spatially resolved imaging would allow probing these states under magnetic inhomogeneity, across interfaces, and in time-dependent fields.

\bibliographystyle{naturemag}
\bibliography{refs}

\twocolumngrid
\section*{Methods}
\methodsec{Device and benchmarks}

The device is a $7\times7$ array of 49 frequency-tunable transmon qubits, each pair of neighbours joined by a tunable coupler. It follows the Zuchongzhi~3.0 architecture~\cite{Gao2025establishing}; the cryogenic set-up and further device details are described in refs~\cite{Wang2026lifting,Tianyan2025cloud}.

Extended Data Fig.~\ref{fig:device}a--f collects the qubit frequencies, coherence times, and the gate and readout benchmarks, and Extended Data Table~\ref{tab:device} summarizes them. The median (mean) relaxation and dephasing times were $T_1=46.9$ ($46.7$)~$\mu\mathrm{s}$ and $T_2^*=3.0$ ($4.5$)~$\mu\mathrm{s}$. Mean Pauli errors were $1.05\times10^{-3}$ for single-qubit gates, $2.90\times10^{-3}$ for $\iswap{1/6}$ gates and $3.72\times10^{-3}$ for $\iswap{1/3}$ gates. The $\iswap{1/6}$ gates were benchmarked on all 84 couplers, the $\iswap{1/3}$ gates on the 42 couplers of groups A and D, since only those two groups execute merged exchange layers (see `Trotter protocol'). Readout assignment errors averaged $P(1|0)=1.4\%$ and $P(0|1)=2.8\%$, that is $2.1\%$ on average.

We calibrated two-qubit gates as in ref.~\cite{Wang2026lifting}. 
Parameter scans yielded the effective inter-qubit coupling and frequency detuning, and we then optimized the pulse parameters to reach the target Weyl coordinates. 
With local phases compensated, the target exchange operation in the single-excitation manifold is $\exp[i\theta(a_i^\dagger a_j+a_j^\dagger a_i)]$, where $\theta=p\pi/2$ for an $\iswap{p}$ gate.

We fixed gate durations at 22~ns for $\iswap{1/6}$ and 28~ns for $\iswap{1/3}$ on every participating coupler. 
Each duration included a 2-ns idle period at the start and an 8-ns idle period at the end, and we optimized the pulse edges coupler by coupler. 
Fixing the durations kept the gates in a parallel group synchronized, and we used the same grouping for calibration and circuit execution. 

We sampled each circuit with $6\times10^4$ shots for rounds 1--8, increasing sampling at greater depth to $4.6\times10^5$ shots at round 15. 
More than 4,000 post-selected outcomes survived at every measured depth (Extended Data Fig.~\ref{fig:device}g,h). 
For the uniform-field circuits, the post-selection rate dropped from $43$--$48\%$ after one round to $1.5$--$2.0\%$ after fifteen rounds, that is, after 46 exchange layers. 

\methodsec{Compilation of the Peierls phases}

\methodsub{Bond conventions and gauge choice}

Horizontal and vertical bonds are oriented along $+x$ and $+y$, respectively. 
For a bond $b:i\to j$ the term $-Je^{i\phi_b}a_j^\dagger a_i$ moves an excitation from $i$ to $j$, and the plaquette flux is $\Phi_p=\phi_{\rm bottom}+\phi_{\rm right}-\phi_{\rm top}-\phi_{\rm left} \pmod{2\pi}$. 
On the open $7\times7$ lattice the 36 fluxes and the 84 bond phases obey $\bm\Phi=A\bm\phi\pmod{2\pi}$, with $A$ the $36\times84$ plaquette--bond incidence matrix (entries $0,\pm1$, rank 36).

For the uniform-field, random-flux and flux-step experiments the bond phases came from
\begin{equation}
\bm\phi=A^+\bm\Phi,
\label{eq:pinv}
\end{equation}
with $A^+$ the Moore--Penrose pseudoinverse. For the real representatives of $\bm\Phi$ chosen here this is the minimum-norm solution of the unwrapped linear system. Every phase was then wrapped to $(-\pi,\pi]$ before the circuit was constructed.

The chiral-transport experiments used the Landau gauge instead, with zero phase on horizontal bonds and $\phi_{(x,y)\to(x,y+1)}=\sum_{x'<x}\Phi_{(x',y)}$ on vertical bonds, which keeps the phase of the horizontal initial bond at zero. The pseudoinverse gauge spreads the phases over both bond directions, whereas the Landau gauge uses vertical bonds only; both reproduce the prescribed fluxes modulo $2\pi$ to numerical precision (Extended Data Fig.~\ref{fig:gauge}a--f).

\methodsub{Gauge covariance of the circuit}

Under the local transformation $G=\exp(i\sum_k\chi_k n_k)$ the hopping operator becomes $Ga_j^\dagger a_iG^\dagger=e^{i(\chi_j-\chi_i)}a_j^\dagger a_i$. Replacing each bond phase by $\phi'_{ij}=\phi_{ij}+\chi_j-\chi_i$ therefore gives
\begin{equation}
U_{\rm T}(\bm\phi')=G\,U_{\rm T}(\bm\phi)\,G^\dagger,
\end{equation}
where $U_{\rm T}$ is the Trotter propagator. The relation holds gate by gate, hence for any layer order.

Site populations do not change when circuit and initial state are transformed together. For a single-site initial state $G$ contributes a global phase only, so the populations are fixed by the plaquette fluxes alone. Numerically, the two gauges agree to a total-variation distance below $10^{-14}$ at every round, for the uniform, random and flux-step landscapes (Extended Data Fig.~\ref{fig:gauge}g--i). As a control, reversing the phases of one coupling group changes the fluxes and yields different populations.

For a two-site initial state the relative phase has to follow the gauge transformation as well. The bonding state on a bond of phase $\phi_{ab}$ is $(|a\rangle+e^{i\phi_{ab}}|b\rangle)/\sqrt2$, which on the horizontal bonds used for chiral transport reduces in the Landau gauge to $(|a\rangle+|b\rangle)/\sqrt2$.

\methodsub{Physical implementation of the phase rotations}

Physical single-qubit $Z$ rotations carry the Peierls phases and cancel calibrated local dynamical phases of the native exchange gates. Rotations about the same axis at the same circuit position are combined by adding angles. A new flux landscape therefore only updates the single-qubit rotations; the calibrated exchange pulses are reused and no entangling gates are added.

Gate calibration and benchmarking used the standard exchange matrix as the reference model, and no independent conditional-phase parameter was fitted. The compilation identities below are derived for the target exchange operation.

\methodsub{Circuit compilation and pulse depth}

The circuit was optimized for the prescribed initial states. Before the bit-flip frames described below were introduced, exchange gates whose two qubits sat in $|00\rangle$ immediately before the gate were removed from the target evolution. Such gates act trivially on the reachable state, so removing them leaves the dynamics of those inputs intact.

We then redistributed the remaining local rotations, using the fact that exchange gates commute with a common $Z$ rotation on their two endpoints. In our schedule the union of two consecutive coupling groups is acyclic, so inside each non-overlapping two-layer block the intermediate rotation angles split into two sets, one commuting with each exchange layer. Within a set the angles are equal on the two endpoints of every bond, but may differ from bond to bond. Rotations moved in this way are placed at the block boundaries.

The remaining pulse depth was reduced with the symmetry
\begin{equation}
(X_iX_j)U_{ij}^{(0)}(X_iX_j)=U_{ij}^{(0)}
\end{equation}
of the zero-phase target exchange operation $U_{ij}^{(0)}=\exp[i\theta(a_i^\dagger a_j+a_j^\dagger a_i)]$. Bit-flip layers were inserted in matched pairs acting on the same set of qubits; every exchange gate in between then had either both endpoints flipped or neither, so the paired flips close the frame without altering the target evolution. Rotation angles are updated as $XZ(\alpha)=Z(-\alpha)X$.

A standalone $Z$ rotation costs four native $\XYP$ pulses when implemented directly; combined with a bit flip it costs two. With $Z(\alpha)=\exp(-i\alpha Z/2)$ the identities read
\begin{equation}
    \begin{aligned}
    \XYP(\varphi)
    &=\exp\!\left[-\frac{i\pi}{4}
    \left(\cos\varphi\,X+\sin\varphi\,Y\right)\right],\\
    Z(\alpha)X
    &=XZ(-\alpha)
    \doteq \bigl[\XYP(\alpha/2)\bigr]^2,
    \end{aligned}
\end{equation}
where $\doteq$ means equality up to a global phase. They hold for the target exchange operations, with the calibrated local phase corrections absorbed into the rotation angles; coherent errors of the native gates are not part of this ideal compilation model.

The compiled bulk schedule carries two $\XYP$ layers per two exchange layers, so for $n$ exchange layers the single-qubit pulse depth drops from roughly $4n$ to $n$. Taking 28~ns for a native single-qubit pulse, the bulk evolution times are
\begin{equation}
    \begin{aligned}
    T_{\rm direct}
    &\simeq n(\tau_2+112\,\mathrm{ns}),\\
    T_{\rm compiled}
    &\simeq n(\tau_2+28\,\mathrm{ns}),
    \end{aligned}
\end{equation}
where $\tau_2$ is the mean exchange-layer duration. For $\tau_2=22$--$28$~ns this is a $60.0$--$62.7\%$ shorter bulk circuit, excluding state preparation, readout and circuit-boundary overheads.

\methodsub{Validation of phase programming}

We checked the programmed landscapes against the Trotterized HH model using the transport exponent, the local interference ratio, and the transmitted population (Fig.~\ref{fig:uniform}c,d and Fig.~\ref{fig:random}f). These three comparisons probe how the programmed fluxes affect interference and transport.

For single-site initial states the bipartite symmetry of the model gives $P_i(t;-\bm\Phi)=P_i(t;+\bm\Phi)$, so populations alone cannot fix the overall field orientation. The bonding and antibonding states of the snake-state experiments supply that check, because the chiral displacement reverses when the field is inverted.

\methodsec{Trotter protocol}

\methodsub{Coupling groups and schedule}

Horizontal bonds belong to A or B according to whether the left site has even or odd $x+y$; vertical bonds belong to C or D according to the parity of the lower site. Each group holds 21 disjoint bonds.

Writing $U_\mu=e^{-iH_\mu\delta t}$ for the evolution on group $\mu$, a forward round applies A, C, B, D and the next round reverses the order, so that each pair of rounds implements
\begin{equation}
    U_{\rm pair}=U_AU_CU_BU_D^2U_BU_CU_A,
\end{equation}
an approximation to evolution over $2\delta t$ with local error $O(\theta^3)$, where $\theta=J\delta t$; at fixed total time the accumulated error is $O(\theta^2)$~\cite{Suzuki1990fractal,Childs2021theory}. The identical D or A layers at a round boundary merge into a single $\iswap{1/3}$ layer. With merging, $k$ rounds give $3k+1$ exchange layers before any input-dependent pruning, and only groups A and D need $\iswap{1/3}$ gates.

\methodsub{Choice of the Trotter step}

We took $\theta=\pi/12$, the native $\iswap{1/6}$ gate, to balance digitization error against circuit depth; smaller steps approximate continuous-time dynamics better but need more rounds at a fixed simulated time.

At $Jt=\pi/2$ the total-variation distances between the Trotterized and the continuous-time site populations are $0.15$, $0.10$ and $0.08$ for $\Phi=0$, $\pi/2$ and $\pi$, and they decrease roughly quadratically with $\theta$ (Extended Data Fig.~\ref{fig:trotter}a,b). An $\iswap{1/8}$ gate would require one-third more rounds at the same simulated time, whereas $\iswap{1/4}$ gives larger digitization errors. Unless stated otherwise, the experimental dynamics are compared with numerical simulations of the Trotterized HH model.

\methodsub{Effect of the layer order}

Updating bonds sequentially can bias the propagation at finite step size. The bias is measured by $V_{C_4}$, the variance of the populations in four $90^\circ$ sectors around the initial site; sites on a dividing axis contribute half their population to each adjacent sector, and the central site contributes one quarter to each.

For a central excitation at zero flux, a fixed layer order gives $V_{C_4}\sim3\times10^{-3}$ after the first few rounds, while the alternating order keeps $V_{C_4}$ below $4\times10^{-4}$ through rounds 1--9 (Extended Data Fig.~\ref{fig:trotter}d).

\methodsub{Fitting window for the transport exponent}

The fit $\mathrm{MSD}\propto t^{\alpha_{\rm eff}}$ uses rounds 2--6 at every flux, that is $Jt\in[\pi/6,\pi/2]$. Leaving out the first round lowers the sensitivity to readout background and early transients, and stopping at round 6 limits the effect of later boundary reflections. Extended Data Fig.~\ref{fig:trotter}c shows how the fitted exponent depends on the window chosen.

The exponent therefore describes spreading over a finite interval and does not determine the asymptotic transport regime. Over the window used, $\alpha_{\rm eff}(\Phi)$ is non-monotonic and reaches its minimum near $\Phi=0.73\pi$.

\methodsub{Trotter error of the chiral displacement}

We compare the chiral displacement $C$ of the two-site initial states between continuous-time and Trotterized HH evolution. At $\Phi=\pi/4$ and $Jt=14\pi/12$ continuous-time evolution gives $C=1.18$ lattice spacings. The $\iswap{1/6}$ protocol ($\theta=\pi/12$) gives $C=1.15$, while the coarser $\iswap{1/3}$ protocol ($\theta=\pi/6$) gives $C=0.44$ (Extended Data Fig.~\ref{fig:chiraltheory}c,d).

\methodsec{Post-selection and readout-error mitigation}

\methodsub{Single-excitation post-selection}

The target evolution conserves the excitation number, and each experiment starts with one excitation. After reading all $L=49$ qubits in the computational basis we keep the bitstrings with $\sum_i m_i=1$ and build the post-selected distribution $p_i^{\rm ps}=N_i/N_{\rm ps}$, where $N_i$ counts the retained outcomes with the excitation on site $i$ and $N_{\rm ps}=\sum_iN_i$.

Inside the single-excitation manifold any single assignment error changes the observed excitation number and the shot is rejected. An accepted shot with the excitation on the wrong site needs a $1\to0$ error on the occupied qubit together with a $0\to1$ error elsewhere, so paired errors of this kind redistribute population inside the retained sample.

\methodsub{Restricted response matrix}

Readout is modelled with independently calibrated assignment errors $\epsilon_i^{0\to1}=P(m_i=1|n_i=0)$ and $\epsilon_i^{1\to0}=P(m_i=0|n_i=1)$. If the errors of different qubits are independent, a true excitation on site $i$ is reported on site $j$ with probability
\begin{equation}
\begin{aligned}
    R_{ii}
    &=(1-\epsilon_i^{1\to0})
      \prod_{k\neq i}(1-\epsilon_k^{0\to1}),\\
    R_{ji}
    &=\epsilon_i^{1\to0}\epsilon_j^{0\to1}
      \prod_{k\neq i,j}(1-\epsilon_k^{0\to1}),
      \qquad j\neq i.
    \end{aligned}
    \label{eq:R}
\end{equation}
The off-diagonal entries are second order in the assignment errors.

Let $s_i=\sum_jR_{ji}$ be the probability that a shot with a true excitation on site $i$ is retained. Conditioned on retention, a normalized single-excitation distribution $\mathbf p$ gives
\begin{equation}
\mathbf p^{\rm ps} = \frac{\mathbf R\mathbf p}{\sum_i s_ip_i}.
\end{equation}
Normalizing each column, $\widetilde R_{ji}=R_{ji}/s_i$, turns this into $\mathbf p^{\rm ps}=\widetilde{\mathbf R}\mathbf q$, where $q_i=s_ip_i/\sum_js_jp_j$ is the acceptance-weighted distribution.

The correction applied here is
\begin{equation}
    \mathbf p^{\rm corr}
    = \widetilde{\mathbf R}^{-1}\mathbf p^{\rm ps}.
\end{equation}
The linear-inversion estimates are retained without clipping negative values, and every observable is computed from the resulting $\mathbf p^{\rm corr}$. This corrects position misassignment within the retained sample but not the reweighting caused by site-dependent acceptance. The column-normalized response matrix $\widetilde{\mathbf R}$ has diagonal elements of $0.96$--$0.99$ and a largest off-diagonal element of about $0.003$. The procedure thus adapts matrix-based readout mitigation to the single-excitation sector~\cite{Bravyi2021mitigating,Nation2021scalable}.

\methodsub{Statistical uncertainty}

Conditioned on $N_{\rm ps}$, the retained counts follow a multinomial
distribution. The estimated covariance of the corresponding site
populations $\mathbf p^{\rm ps}$ is
\begin{equation}
    \boldsymbol\Sigma_{\rm ps}
    =\frac{
    \operatorname{diag}(\mathbf p^{\rm ps})
    -\mathbf p^{\rm ps}(\mathbf p^{\rm ps})^T
    }{N_{\rm ps}}.
\end{equation}
With the calibrated response matrix held fixed, linear propagation through the inversion gives
\begin{equation}
    \boldsymbol\Sigma_{\rm lin}
    =\widetilde{\mathbf R}^{-1}
     \boldsymbol\Sigma_{\rm ps}
     (\widetilde{\mathbf R}^{-1})^T.
\end{equation}
For a linear observable $O=\sum_i o_i p_i^{\rm corr}$ the variance estimate is $\sigma_O^2\simeq\mathbf o^T\boldsymbol\Sigma_{\rm lin}\mathbf o$. For nonlinear observables, we propagate the covariance using the gradient of the observable.

The four circuits that enter the chiral displacement are sampled independently, so
\begin{equation}
    \sigma_C^2 =\frac{1}{16} \sum_{s=\pm}\sum_{\eta=B,A}\sigma_{X_{s,\eta}}^2.
\end{equation}
Error bars derived from populations are one standard error from shot statistics; for disorder averages they are conditional on the sampled realizations.
In Fig.~\ref{fig:random}b at $W>0$ the uncertainty of the mean exponent, propagated from the individual fits, is combined in quadrature with the standard error across realizations.
At $W=0$ the error bar is the fit uncertainty reported in Fig.~\ref{fig:uniform}c.
Variation between finite numbers of realizations is assessed below.

\methodsub{Validity and limitations}

The restricted response model accounts for assignment errors inside the single-excitation manifold, but states outside it can also pass post-selection. A vacuum state is accepted when exactly one qubit is falsely assigned to $|1\rangle$, with probability
\begin{equation}
    P_{\rm acc}^{\rm vac}
    =\sum_{j=1}^{L}\epsilon_j^{0\to1}
     \prod_{k\neq j}(1-\epsilon_k^{0\to1}),
\end{equation}
and a two-excitation state enters through a $1\to0$ error.

Contamination of this kind survives the restricted correction and can leave residual population whose spatial profile follows the assignment errors. Correlated readout errors and measurement crosstalk also fall outside the independent-error model. The corrected distributions therefore describe the post-selected sample, with the acceptance bias noted above; they are not a reconstruction of the full many-qubit state.

\methodsub{Post-selection budget and sample contamination}

Errors that change the excitation number remove shots from the post-selected sample. With 48 qubits in $|0\rangle$, a Pauli error carrying an $X$ or $Y$ component almost always creates an excitation. Per round the three exchange layers contribute $63\times2.9\times10^{-3}\times12/15\approx0.15$ such errors and the three XY2P layers $147\times1.05\times10^{-3}\times2/3\approx0.10$, giving $r\approx0.25$ in total. For a single excitation on site $i$, the probability of reporting it at the correct site with no false positives is $R_{ii}=(1-\epsilon_i^{1\to0})\prod_{j\ne i}(1-\epsilon_j^{0\to1})
\approx 0.49$ for typical readout errors. A Poisson model with $r\approx0.25$ number-changing errors per round, readout errors and $T_1$ decay reproduces the measured post-selection rate, which falls approximately as $e^{-0.235k}$ over the first fifteen rounds (Extended Data Fig.~\ref{fig:budget}a).

Background events can arise through several processes, including a two-excitation state in which the original excitation is misread as 0 while the error-created excitation is correctly detected, a two-excitation state whose original excitation decays while the error-created one survives, and the vacuum state with exactly one false positive. Extended Data Fig.~\ref{fig:budget}b shows the measured fractions of shots with zero, one, two or at least three detected excitations at each round, averaged over the uniform fields $\Phi=0$, $\pi/2$ and $\pi$. Fitting the model, including readout errors, to these fractions gives $0.28$ number-changing errors per round and an effective false-positive probability of $1.1\%$ per qubit (root-mean-square deviation $0.016$, compared with $0.020$ for the benchmarked rates). Before readout errors are applied, the fitted model predicts that $33\%$, $36\%$ and $30\%$ of shots contain one, two and at least three excitations, respectively, after four rounds. After twelve rounds, the corresponding fractions are $4\%$, $13\%$ and $82\%$. The model separates accepted events into detection of the original excitation ($A_{\rm true}$), background events ($A_{\rm rand}$), and single-excitation readout misassignments treated by the readout correction ($A_{\rm corr}$). We approximate the effective background fraction as $f_{\rm bg}=A_{\rm rand}/(A_{\rm true}+A_{\rm rand})$, excluding $A_{\rm corr}$ from this normalization. Under this model, the estimated effective background fraction is $5\%$ after four rounds, $12\%$ after eight and $20\%$ after twelve. We use the background fractions estimated from the uniform-field circuits to predict the reduction in chiral displacement shown in Extended Data Fig.~\ref{fig:budget}c.

If one background distribution contributes the same fraction $f$ to all four measurements, it rescales $C$, $V$ and $D$ by $1-f$ and leaves $V=D=0$ for an otherwise
ideal signal. For a uniform background,
\[
\mathrm{MSD}_{\rm obs}
=(1-f)\mathrm{MSD}_{\rm sig}+8f,
\qquad
\mathcal T_{\rm obs}
=(1-f)\mathcal T_{\rm sig}+\frac{3f}{7}.
\] At round 12, the inferred background fraction of $20\%$ reduces the predicted $C(\pi/4,k)$ to about $80\%$ of its ideal-circuit value, slightly below the $84$--$89\%$ measured at late rounds (Extended Data Fig.~\ref{fig:budget}c); the predicted reduction of $\alpha_{\rm eff}$ at $\Phi=0$ ($1.80$ against the measured $1.91$) and the predicted excess of $\mathcal T$ at $\Phi=\pi$ ($0.18$ against $0.155$) are also a little too large. 

\methodsec{Theory of the chiral measurement}

\methodsub{Symmetry relation}

For an oriented bond $b:i\to j$ define $h_b(\phi_b)=e^{i\phi_b}a_j^\dagger a_i+\mathrm{h.c.}$ and $S=\operatorname{diag}[(-1)^{x+y}]$. The hopping operators obey $h_b(\phi_b)^*=h_b(-\phi_b)$ and $Sh_bS=-h_b$, so every exchange layer satisfies
\begin{equation}
L(-\boldsymbol\phi) = S L(\boldsymbol\phi)^*S.
\end{equation}
This holds for arbitrary real angles that may differ from bond to bond, and it survives multiplication in any layer order. Hence $U(-\boldsymbol\phi)=S U(\boldsymbol\phi)^*S$, and
\begin{equation}
    P_i(-\boldsymbol\Phi,\psi) = P_i(+\boldsymbol\Phi,S\psi^*).
    \label{eq:theorem}
\end{equation}

For a single-site initial state $S\psi^*=\pm\psi$, so the populations are insensitive to a global flux reversal (Extended Data Fig.~\ref{fig:chiraltheory}a). In the Landau gauge the horizontal initial bond has zero phase; on that bond $S\psi^*$ exchanges the bonding and antibonding states, while $(|a\rangle\pm i|b\rangle)/\sqrt2$ is invariant up to a global phase. The centroids of the target circuit therefore obey $X_{-,B}=X_{+,A}$ and $X_{-,A}=X_{+,B}$.

The same relations hold for noise channels covariant under $\rho\mapsto S\rho^*S$, as long as the channels are identical in the paired circuits. Independent $Z$ dephasing and amplitude damping satisfy this condition in the logical evolution, and a common classical readout map preserves the equality of the paired distributions. Static on-site detunings and additive Peierls-phase offsets change sign under the transformation, and can break the relations whenever they are fixed between circuits.

\methodsub{Four-circuit decomposition and error cancellation}

The centroids for orientation $s=\pm$ and state $\eta=B,A$ are combined as
\begin{equation}
    \begin{aligned}
    M&=\tfrac14(X_{+,B}+X_{-,B}+X_{+,A}+X_{-,A}),\\
    C&=\tfrac14(X_{+,B}-X_{-,B}-X_{+,A}+X_{-,A}),\\
    V&=\tfrac14(X_{+,B}+X_{-,B}-X_{+,A}-X_{-,A}),\\
    D&=\tfrac14(X_{+,B}-X_{-,B}+X_{+,A}-X_{-,A}).
    \end{aligned}
\end{equation}
Equation~(\ref{eq:theorem}) gives $V=D=0$ for the target circuit and for noise obeying the covariance condition above.

Consider a static perturbation $\varepsilon$ that is shared by all four circuits and reversed by the symmetry,
\begin{equation}
X_{-s,\bar\eta}(\varepsilon) = X_{s,\eta}(-\varepsilon),
\end{equation}
where $\bar\eta$ is the partner state of $\eta$. Then $M$ and $C$ are even in $\varepsilon$ and $V$ and $D$ are odd:
\begin{equation}
    \begin{aligned}
    C(\varepsilon)&=C(0)+O(\varepsilon^2),\\
    M(\varepsilon)&=M(0)+O(\varepsilon^2),\\
    V(\varepsilon)&=O(\varepsilon),\\
    D(\varepsilon)&=O(\varepsilon).
    \end{aligned}
\end{equation}
The residuals therefore probe symmetry-breaking errors that, under this shared-error model, do not affect $C$ at first order. The cancellation does not carry over to errors that differ between the four circuits.

Both measured residuals stay below $0.045$ in magnitude (Fig.~\ref{fig:chiral}e). Their error-weighted means over fourteen rounds are $V=-0.003\pm0.001$ and $D=-0.005\pm0.001$, and at rounds 8--13 both reach $-0.03$ to $-0.04$. These departures measure a residual asymmetry between the paired measurements; on their own they do not bound the second-order error in $C$.

\methodsub{Symmetry-preserving errors}

Errors that are invariant under the symmetry, such as decoherence, a common error of the exchange angle or the post-selection background, satisfy $X_{-s,\bar\eta}(\varepsilon)=X_{s,\eta}(\varepsilon)$. They leave $V=D=0$ while changing $C$ at first order. To model dephasing, we apply an independent dephasing channel to each qubit after every exchange layer. Each channel attenuates single-qubit coherence by a factor $e^{-\gamma}$. Since the single-excitation states $|i\rangle$ and $|j\rangle$ differ on two qubits for $i\ne j$, the off-diagonal element $\rho_{ij}$ acquires a factor $e^{-2\gamma}$ per layer. This dephasing preserves the symmetry relations. Fitted to $C(\pi/4,k)$ without the post-selection background, a single value $\gamma=0.004$ per layer reproduces the saturation of the measured displacement at $84$--$89\%$ of the ideal value (Extended Data Fig.~\ref{fig:budget}c), the chiral velocity ($0.44$ against $0.52\,Ja/\hbar$) and $\alpha_{\rm eff}(\Phi=0)$ ($1.91$, as measured). For a layer duration of 50--56~ns, the background-free fit $\gamma=0.004$ corresponds to an effective single-qubit dephasing time of about $13\,\mu\mathrm{s}$ during the circuit, longer than the Ramsey dephasing time $T_2^*$. 

\methodsub{Phase scan}

In the experimental gauge the initial state is $|\psi_\beta\rangle=(|a\rangle+e^{i\beta}|b\rangle)/\sqrt{2}$. Its density matrix is linear in $\cos\beta$ and $\sin\beta$, and so is the centroid at a fixed round under the ideal circuit. Since $S|\psi_\beta^*\rangle$ equals $|\psi_{\pi-\beta}\rangle$ up to a global phase,
\begin{equation}
    X_-(\beta)=X_+(\pi-\beta).
\end{equation}
The ideal response can therefore be written as $X_s^{(0)}(\beta)=x_0+sF\cos\beta+G\sin\beta$ with $s=\pm1$, which gives
\begin{equation}
    \frac{X_+^{(0)}(\beta)-X_-^{(0)}(\beta)}{2}
    =F\cos\beta.
\end{equation}
Here $F$ equals the ideal four-circuit chiral displacement $C$ obtained from the bonding and antibonding preparations, while the sine component is even under field reversal (Extended Data Fig.~\ref{fig:chiraltheory}b).

At round 10 and $\Phi=\pi/4$ the sixteen measured centroids are fitted jointly to
\begin{equation}
    X(\beta,s)=a_0+a_1s
    +(b+sf)\cos\beta+(g+sh)\sin\beta.
\end{equation}
The ideal symmetry requires $a_1=b=h=0$. The weighted fit gives $f=0.987\pm0.008$ and $g=0.786\pm0.010$, with $a_1=-0.022\pm0.006$, $b=-0.006\pm0.008$ and $h=-0.064\pm0.010$. The fitted chiral amplitude agrees with the independently extracted four-circuit value $C(10)=0.99\pm0.01$.

Preparation-phase offsets can be assessed inside an otherwise ideal model. For a common offset $\beta_{\mathrm{actual}}=\beta-\delta$ the fitted coefficients satisfy
\[
    h=f\tan\delta,\qquad b=-g\tan\delta.
\]
The measured $b$ and $g$ then give the small-offset estimate $\delta\simeq0.008\pm0.010$ rad, which does not account for the measured $h$. A common preparation-phase offset alone therefore leaves the observed symmetry breaking unexplained.

An orientation-dependent offset
$\beta_{\mathrm{actual}}=\beta-s\delta$ instead gives
\[
    f=F\cos\delta-G\sin\delta,\qquad
    g=G\cos\delta+F\sin\delta,
\]
while preserving $a_1=b=h=0$. It can shift the extracted chiral amplitude by $-G\delta+O(\delta^2)$ without producing nonzero
four-circuit residuals $V$ or $D$. The cancellation derived for shared static perturbations does not in general protect against such
orientation-dependent preparation errors. An offset of this kind, however, moves $f$ and $g$ in opposite directions, whereas both measured amplitudes are approximately $0.88$ times their ideal values ($f=0.99$ against $1.12$ and $g=0.79$ against $0.89$). Such a common reduction can arise from symmetry-preserving errors, including the post-selection background or the dephasing discussed above.

The fit gives $\chi^2=52$ for 10 degrees of freedom, so the data deviate from the six-parameter first-harmonic model by more than the estimated shot-noise uncertainties. Phase-dependent preparation imperfections and post-selection probabilities can both contribute. The fit is therefore used to characterize the dominant phase dependence; the quoted uncertainties do not include these systematic effects, and the nonzero $h$ alone does not distinguish errors in state preparation, evolution and measurement.

\methodsub{State preparation}

Starting from $|0\cdots0\rangle$, qubit $a=(3,3)$ is excited with an $X$ gate, a separately calibrated $\iswap{1/2}$ gate is applied on the bond to $b=(4,3)$, and the relative phase is set by an $R_z$ rotation on $b$ of angle $\phi_{ab}+\beta-\pi/2$. The $\iswap{1/2}$ gate has exchange angle $\pi/4$ and populates the two sites equally; it was calibrated independently of the $\iswap{1/6}$ gates used in the Trotter steps.

In the phase scan, $\beta$ takes eight values for each orientation. For the distance scan the same preparation is applied to the bond $(3,d)$--$(4,d)$ on rows $d=3,2,1,0$, that is on the reversal row and one, two and three rows below it (Fig.~\ref{fig:chiral}f).

\methodsub{Strip bands and initial-state overlaps}

Bands are calculated for a seven-row strip, infinite along $x$, carrying a $+\Phi\,|\,-\Phi$ reversal (Extended Data Fig.~\ref{fig:spectra}). In a gauge with real vertical hopping and horizontal phases satisfying $\phi_x(y+1)=\phi_x(y)-\Phi_{\rm row}(y)$, the Bloch Hamiltonian is
\begin{equation}
    \begin{aligned}
    H(k)_{yy}&=-2J\cos[k-\phi_x(y)],\\
    H(k)_{y,y+1}&=H(k)_{y+1,y}=-J
    \end{aligned}
\end{equation}
for dimensionless quasimomentum $k$. The reversal produces a piecewise-linear vector potential and interface modes tied to snake-state motion~\cite{Muller1992electron,Reijniers2000snake}. States with a weight above 0.5 on the three rows around the line are classed as interface-dominated (Extended Data Fig.~\ref{fig:spectra}a,b). The overlaps of the initial states with the strip eigenstates determine the weighting of their group velocities.

For the initial bond on row $y=3$ the overlap with band $n$ is
\begin{equation}
w_n(k)\propto
\left|1+e^{i[\phi_x(3)+\beta-k]}\right|^2
|u_{n,3}(k)|^2,
\end{equation}
where $u_{n,3}(k)$ is the eigenvector component on that row. At $\Phi=\pi/4$ the bonding state has $72\%$ of its spectral weight at $E<0$ and $75\%$ on interface-dominated branches, with an overlap-weighted velocity of $+0.55\,Ja/\hbar$; the antibonding state is mirrored to positive energies with velocity $-0.55\,Ja/\hbar$ (Extended Data Fig.~\ref{fig:spectra}c--e).

Fitted over rounds 4--10, the chiral velocity is $0.58\,Ja/\hbar$ for continuous-time evolution on the open $7\times7$ array and $0.52\,Ja/\hbar$ for the Trotterized circuit. The overlap-weighted group velocity of the strip, $0.55\,Ja/\hbar$, is a different estimator and agrees with the open-array value to within $5\%$. On the open $7\times7$ lattice the bonding and antibonding states carry $74\%$ and $26\%$ of their weight at $E<0$ (Extended Data Fig.~\ref{fig:spectra}f).

\methodsub{Flux and distance dependence}

For the Trotterized model the chiral displacement after ten rounds peaks near $\Phi=0.23\pi$ and vanishes at $\Phi=0$ and $\pi$. The corresponding magnetic length, $\ell_B=a/\sqrt{|\Phi|}\approx1.2a$, is comparable to the lattice spacing.

The displacement depends strongly on how far the initial bond lies from the reversal line. At the outer edge its sign reverses, as expected for skipping-orbit transport (Fig.~\ref{fig:chiral}f). The ratio of measured to ideal displacement is $1.05$, $0.83$ and $0.84$ at the reversal row and $0.98$, $0.89$ and $0.81$ at the edge after 4, 8 and 12 rounds; one row below the line it is $0.81$ and $0.90$ after 8 and 12 rounds. The absolute sign depends on the coordinate and hopping-phase conventions; the reversals under field inversion and state interchange do not.

\methodsec{Numerical methods}

Since the target HH Hamiltonian conserves the excitation number, the single-excitation dynamics is a 49-dimensional problem. States are propagated with the exact exchange-layer unitaries of the Trotterized model, or with $e^{-iHt}$ for the continuous-time comparisons. For the random-flux experiments at $W>0$ the model is evaluated on the same 20 realizations as the experiment (Fig.~\ref{fig:random}b,c); $W=0$ uses the uniform zero-field model. To estimate the sampling error of such 20-realization averages we simulated 1,000 realizations and evaluated $\alpha_{\rm eff}$ and $\langle\mathrm{PR}\rangle$ on random subsets of $N$ realizations (Extended Data Fig.~\ref{fig:randomtheory}c,d). At $W=\pi$ the standard deviation over 20-realization subsets, the error bars at $N=20$ in Extended Data Fig.~\ref{fig:randomtheory}c,d, is $0.019$ for $\alpha_{\rm eff}$ and $0.9$ for the participation ratio after six rounds. Large-lattice simulations use a $61\times61$ lattice, Krylov propagation and 24 realizations per $W$ (Extended Data Fig.~\ref{fig:randomtheory}a,b). The experimental comparisons in Figs.~2--4 use the ideal
Trotterized HH model. Every model curve is recomputed by the figure scripts from the circuit definition and checked against the stored values.

\noindent\textbf{Acknowledgements:}
This work was supported by the Quantum Science and Technology-National Science and Technology Major Project 2025ZD0300604. 
W.-Y.Z.~acknowledges support from NSFC under Grant No. 12404323 and No. 12674336, the Fundamental Research Funds for the Central Universities under Grant No. WK2030000104.

\clearpage
\onecolumngrid
\setcounter{figure}{0}
\renewcommand*{\theHfigure}{ED.\arabic{figure}}
\renewcommand{\thefigure}{\arabic{figure}}
\renewcommand{\figurename}{Extended Data Figure}
\setcounter{table}{0}
\renewcommand{\thetable}{\arabic{table}}
\renewcommand{\tablename}{Extended Data Table}

\begin{figure}[p]
    \centering
    \includegraphics[width=0.92\textwidth,height=0.8\textheight,keepaspectratio]{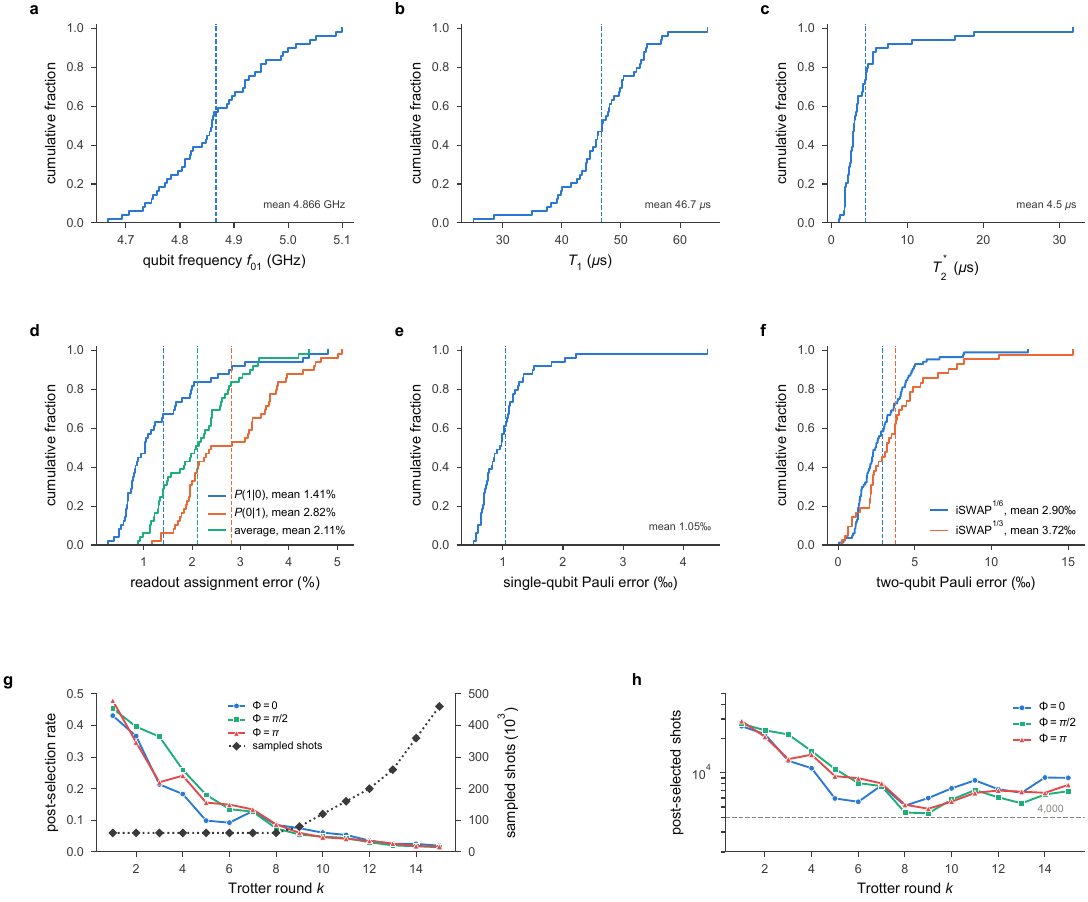}
    \caption{\textbf{Device benchmarks and post-selection statistics.}
    \textbf{a--c}, Cumulative distributions over the 49 qubits of the idle frequencies $f_{01}$ (\textbf{a}), the relaxation times $T_1$ (\textbf{b}) and the dephasing times $T_2^*$ (\textbf{c}).
    \textbf{d}, Cumulative distributions of the readout assignment errors $P(1|0)$, $P(0|1)$ and their average; $P(m|n)$ is the probability of assigning outcome $m$ to a qubit prepared in $|n\rangle$.
    \textbf{e}, Cumulative distribution of the single-qubit Pauli errors of the 49 qubits.
    \textbf{f}, Cumulative distributions of the Pauli errors of the native $\iswap{1/6}$ gates, on all 84 couplers, and of the merged $\iswap{1/3}$ gates, on the 42 couplers of groups A and D. Vertical dashed lines in \textbf{a--f} mark means (in \textbf{d} one per error type, in \textbf{f} one per gate; the values are given in the panels).
    \textbf{g}, Post-selection rate (left axis) against Trotter round $k$ ($Jt=k\pi/12$) for the uniform fluxes $\Phi=0$, $\pi/2$ and $\pi$; the dotted curve gives total shots per circuit on the right axis. Post-selection keeps only outcomes with exactly one excitation across the array.
    \textbf{h}, Post-selected shots per circuit for the data in \textbf{g}; the horizontal dashed line marks 4,000 shots.}
    \label{fig:device}
\end{figure}
\clearpage

\begin{figure}[p]
    \centering
    \includegraphics[width=0.92\textwidth,height=0.8\textheight,keepaspectratio]{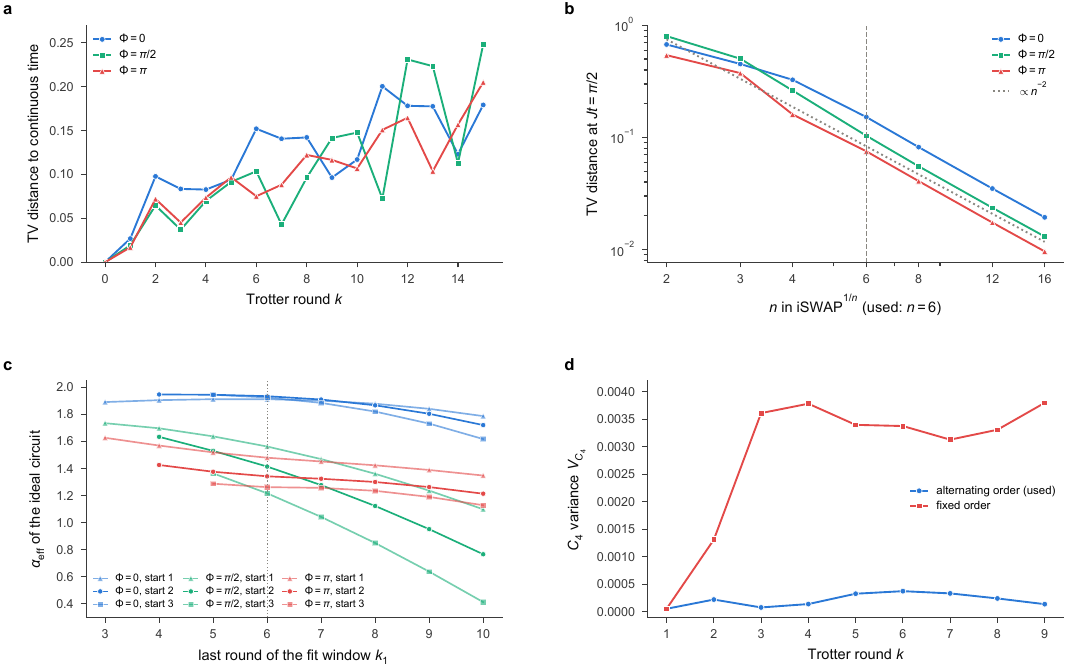}
    \caption{\textbf{The Trotter protocol (simulation).}
    \textbf{a}, Total-variation distance between the site populations of the $\iswap{1/6}$ circuit and of the continuous-time evolution at the same times, for $\Phi=0$, $\pi/2$ and $\pi$.
    \textbf{b}, The same distance at $Jt=\pi/2$ against $n$ in $\iswap{1/n}$; the dotted line is $\propto n^{-2}$.
    \textbf{c}, Effective exponent of the ideal circuit against the fit window (first round 1, 2 or 3; last round $k_1$); the dotted line marks the window used here (rounds 2--6).
    \textbf{d}, Fourfold-symmetry variance $V_{C_4}$ of the $\Phi=0$ distribution, for the alternating schedule and for a fixed order of the four layers.}
    \label{fig:trotter}
\end{figure}
\clearpage

\begin{figure}[p]
    \centering
    \includegraphics[width=0.92\textwidth,height=0.8\textheight,keepaspectratio]{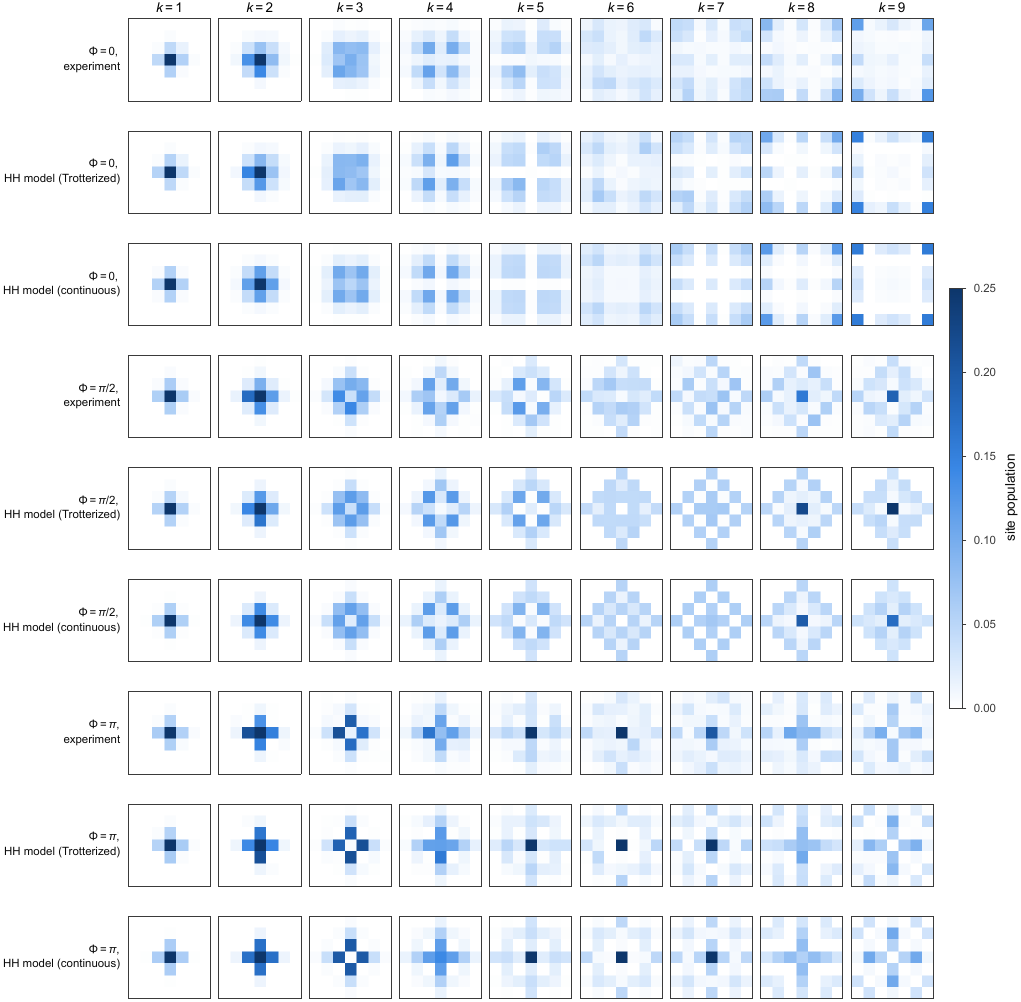}
    \caption{\textbf{Site populations in a uniform synthetic field: experiment and theory.}
    Measured site populations after Trotter rounds $k=1$--$9$ ($Jt=k\pi/12$) at $\Phi=0$, $\pi/2$ and $\pi$ (rows labelled `experiment'; the data are those of Fig.~\ref{fig:uniform}a), each followed by the exact simulation of the Trotterized HH circuit and by the continuous-time evolution of the HH Hamiltonian at the same times. Every panel uses the same colour scale (site population, saturating at 0.25). The digital circuit reproduces the continuous-time dynamics: ballistic expansion at $\Phi=0$, ring-like spreading with partial refocusing at $\Phi=\pi/2$, and axial propagation at $\Phi=\pi$. The measured maps follow the same patterns, with some loss of contrast at the latest rounds.}
    \label{fig:density}
\end{figure}
\clearpage

\begin{figure}[p]
    \centering
    \includegraphics[width=0.92\textwidth,height=0.8\textheight,keepaspectratio]{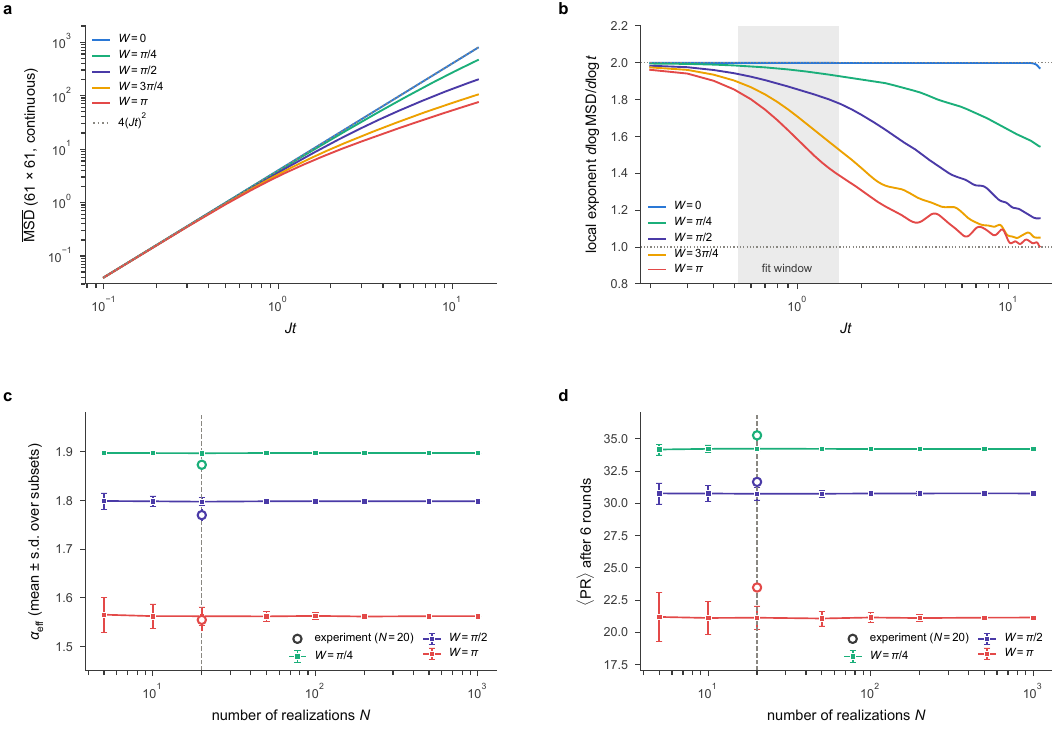}
    \caption{\textbf{Random gauge fields: long times and the number of disorder realizations (simulation).}
    \textbf{a}, Disorder-averaged MSD in continuous time on a $61\times61$ lattice, for five disorder strengths (24 realizations per $W$).
    \textbf{b}, Local exponent $d\log{\rm MSD}/d\log t$ of the curves in \textbf{a}; at $W=\pi$ it approaches 1 at long times. The shaded band is the fit window of the experiment (rounds 2--6).
    \textbf{c}, $\alpha_{\rm eff}$ against the number of realizations $N$ (mean and standard deviation of the $N$-realization average over random subsets of 1,000 realizations of the $7\times7$ Trotter circuit), for three disorder strengths, together with the measured values at $N=20$ (open circles).
    \textbf{d}, The same for the participation ratio after six rounds.}
    \label{fig:randomtheory}
\end{figure}
\clearpage

\begin{figure}[p]
    \centering
    \includegraphics[width=0.92\textwidth,height=0.8\textheight,keepaspectratio]{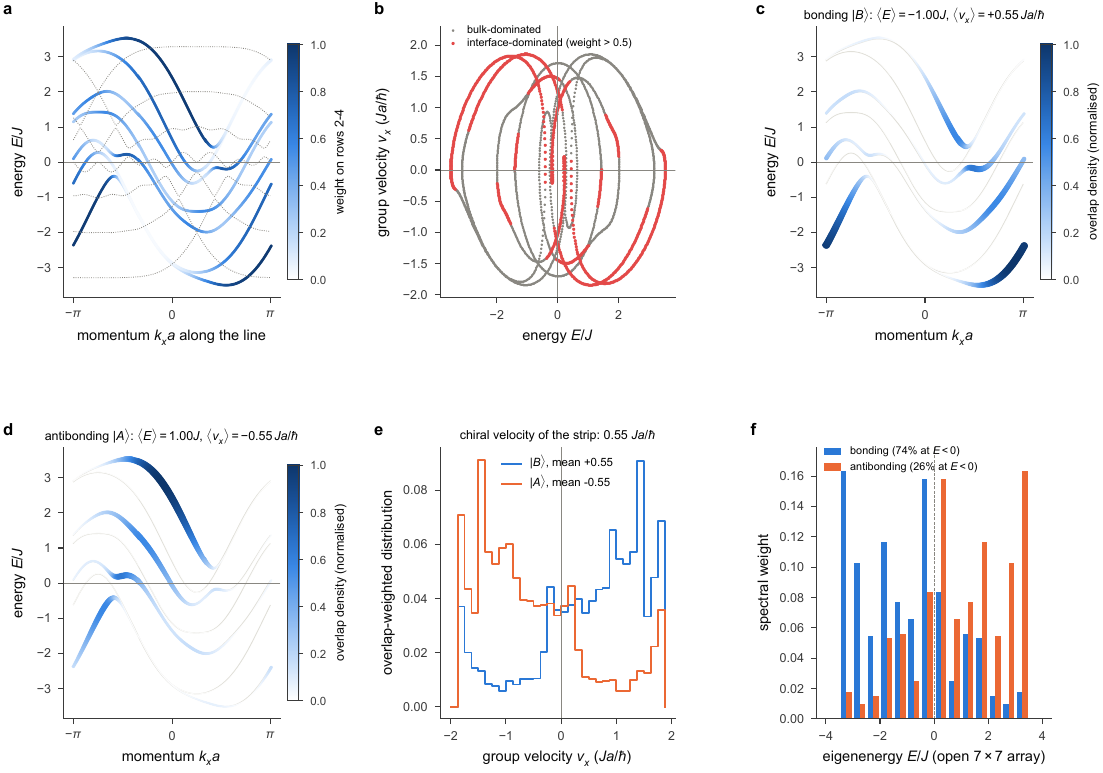}
    \caption{\textbf{Band structure of the field-reversal strip and the initial states (theory).}
    \textbf{a}, Bands of a seven-row strip with the $+\Phi\,|\,-\Phi$ texture ($\Phi=\pi/4$) against momentum along the reversal line; colour gives the weight on the three rows around the line, and the dotted curves are the bands of the uniform field.
    \textbf{b}, Group velocity of the strip eigenstates against energy. Red points mark interface-dominated states (weight $>0.5$ on the three rows centred on the reversal line); grey points are the remaining states.
    \textbf{c}, \textbf{d}, Overlap density of the bonding (\textbf{c}) and antibonding (\textbf{d}) initial state with the strip eigenstates (marker size and colour); the titles give the mean energy and the mean group velocity of each state.
    \textbf{e}, Distribution of the group velocity weighted by the overlap, for both states.
    \textbf{f}, Spectral weights of the two states on the eigenstates of the open $7\times7$ array (bins of $0.5J$).}
    \label{fig:spectra}
\end{figure}
\clearpage

\begin{figure}[p]
    \centering
    \includegraphics[width=0.92\textwidth,height=0.8\textheight,keepaspectratio]{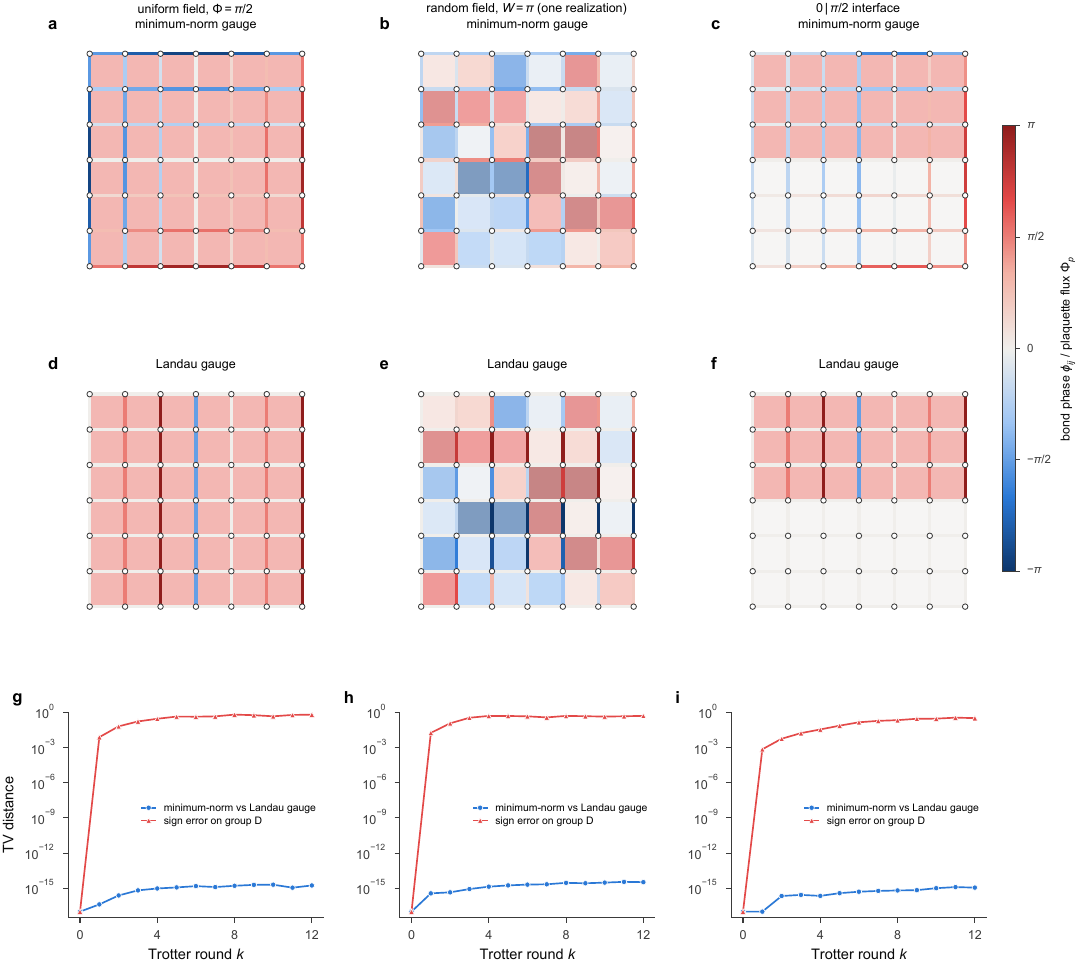}
    \caption{\textbf{Gauge choices and gauge invariance.}
    \textbf{a--c}, Bond phases in the minimum-norm gauge of equation~(\ref{eq:pinv}) for a uniform field with $\Phi=\pi/2$ (\textbf{a}), a random field with $W=\pi$ (one realization; \textbf{b}) and a $0\,|\,\pi/2$ interface (\textbf{c}). Bond colour gives the Peierls phase, plaquette colour the enclosed flux.
    \textbf{d--f}, Landau-gauge implementations of the same landscapes, with phases on vertical bonds only.
    \textbf{g--i}, Total-variation distance between the site populations of the minimum-norm and the Landau gauge (blue) for the landscapes of \textbf{a--c}. Red curves compare the Landau-gauge result with a control in which only the Peierls phases on the bonds of coupling group D are reversed, all other bond phases unchanged. All populations come from numerical simulations of the Trotterized HH model with single-site initial states. The two gauges agree to numerical precision, with total-variation distances below $10^{-14}$.}
    \label{fig:gauge}
\end{figure}
\clearpage

\begin{figure}[p]
    \centering
    \includegraphics[width=0.92\textwidth,height=0.8\textheight,keepaspectratio]{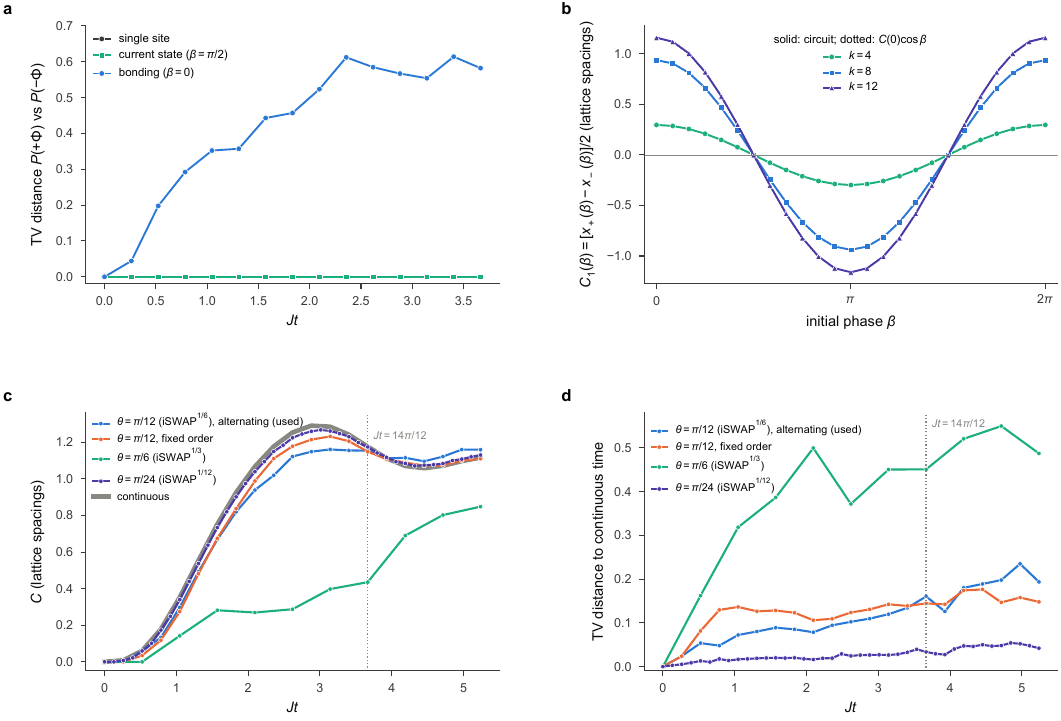}
    \caption{\textbf{Theory of the chiral measurement (simulation).}
    All panels use $\Phi=\pi/4$.
    \textbf{a}, Total-variation distance between the site distributions $P(+\Phi,\psi)$ and $P(-\Phi,\psi)$ for the same initial state $\psi$: a single-site state, a current-carrying state ($\beta=\pi/2$) and a bonding state ($\beta=0$).
    \textbf{b}, Chiral component of the phase scan, $\tfrac12[x_+(\beta)-x_-(\beta)]$, after 4, 8 and 12 rounds (solid curves); it is an exact cosine, $C_1(\beta)=C(0)\cos\beta$ (dotted), at every round.
    \textbf{c}, Trotter error of $C(t)$: continuous time against digital schedules with different step sizes and layer orders; the dotted line marks $Jt=14\pi/12$.
    \textbf{d}, Total-variation distance of the $(+\Phi\,|\,-\Phi,\,B)$ distribution to continuous time, for the same schedules.}
    \label{fig:chiraltheory}
\end{figure}
\clearpage

\begin{figure}[p]
    \centering
    \includegraphics[width=0.92\textwidth,height=0.8\textheight,keepaspectratio]{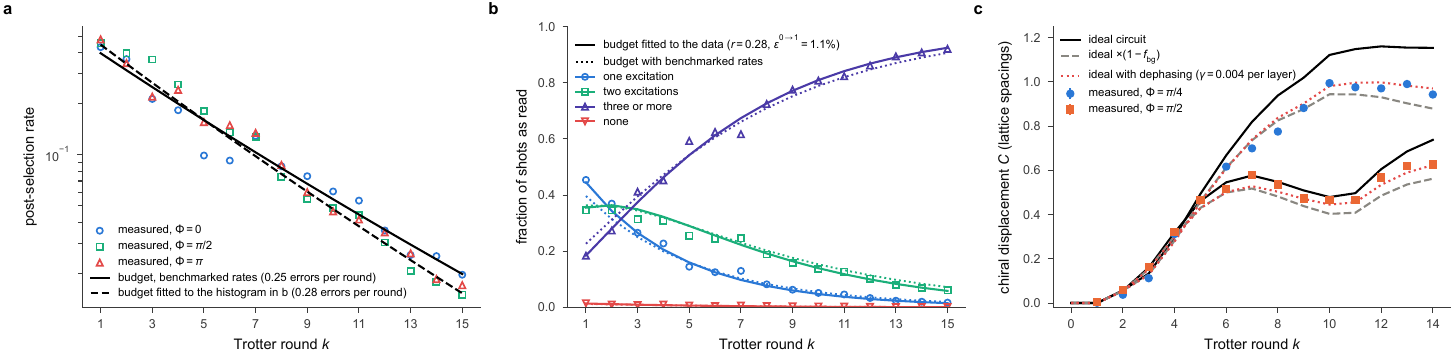}
    \caption{\textbf{Post-selection budget and sample contamination.}
    \textbf{a}, Post-selection rate against Trotter round for the uniform fluxes $\Phi=0$, $\pi/2$ and $\pi$ (symbols), compared with a Poisson budget of number-changing gate errors, readout errors and $T_1$ decay (Methods). Solid curve, benchmarked error rates ($0.25$ per round); dashed curve, rate fitted to the excitation-number statistics in \textbf{b} ($0.28$ per round).
    \textbf{b}, Fractions of shots read with one, two, three or more and no excitation after each round. Symbols are measured for the uniform-field circuits (mean over $\Phi=0$, $\pi/2$ and $\pi$); lines give the budget passed through the readout errors, with the benchmarked rates (dotted) and with the error-creation and false-positive rates fitted to the data (solid; $0.28$ per round and $1.1\%$ per qubit).
    \textbf{c}, Chiral displacement at $\Phi=\pi/4$ and $\pi/2$: measured (symbols), ideal circuit (solid), ideal circuit scaled by $1-f_{\rm bg}$ (dashed), and ideal circuit with dephasing of $\gamma=0.004$ per exchange layer (dotted; `Symmetry-preserving errors'). Here, $f_{\rm bg}$ is the effective background fraction estimated from the error-budget model constrained by the excitation-number statistics in \textbf{b} (Methods).}
    \label{fig:budget}
\end{figure}

\begin{table}[p]
    \centering
    \caption{\textbf{Device parameters and gate benchmarks.} Summary statistics of the distributions in Extended Data Fig.~\ref{fig:device}a--f; $n$ is the number of qubits or couplers. The $\iswap{1/3}$ gates are the merged boundary layers of coupling groups A and D. Gate durations, including idle periods: 22~ns ($\iswap{1/6}$), 28~ns ($\iswap{1/3}$) and 28~ns (XY2P); the readout duration is 800~ns.}
    \label{tab:device}
    \small
    \begin{tabular}{lccccc}
\toprule
Quantity & $n$ & Mean & Median & Min. & Max. \\
\midrule
Qubit frequency $f_{01}$ (GHz) & 49 & 4.866 & 4.860 & 4.667 & 5.100 \\
Relaxation time $T_1$ ($\mu$s) & 49 & 46.7 & 46.9 & 25.1 & 64.7 \\
Dephasing time $T_2^*$ ($\mu$s) & 49 & 4.5 & 3.0 & 1.0 & 31.8 \\
Single-qubit gate Pauli error (\textperthousand) & 49 & 1.05 & 0.95 & 0.51 & 4.40 \\
$\mathrm{iSWAP}^{1/6}$ Pauli error (\textperthousand) & 84 & 2.90 & 2.45 & 0.06 & 12.39 \\
$\mathrm{iSWAP}^{1/3}$ Pauli error (\textperthousand) & 42 & 3.72 & 3.20 & 0.26 & 15.32 \\
Readout error $P(1|0)$ (\%) & 49 & 1.41 & 1.03 & 0.27 & 4.82 \\
Readout error $P(0|1)$ (\%) & 49 & 2.82 & 2.40 & 1.17 & 5.10 \\
Average readout assignment error (\%) & 49 & 2.11 & 2.08 & 0.89 & 4.43 \\
\bottomrule
\end{tabular}

\end{table}

\end{document}